\documentclass[manuscript, screen]{acmart}

\usepackage{svg}
\usepackage{subfig}
\usepackage{tcolorbox}

\usepackage[normalem]{ulem}

\usepackage{longtable}

\usepackage{xcolor,colortbl}

\usepackage{algorithmic}
\usepackage{graphicx}
\usepackage{textcomp}
\usepackage{xcolor}
\usepackage{mdframed}

\usepackage{hyperref}
\usepackage{glossaries}
\usepackage{graphicx}
\usepackage{listings}
\usepackage{csquotes}
\usepackage[inline]{enumitem}
\usepackage{dblfloatfix}

\lstdefinelanguage{wfm}[]{Java}{
  morekeywords={process, task, subprocess, adhoc, transaction, gateway, event, start, end, boundary, data, store, split, merge, lane}
}
\lstdefinelanguage{tagging}[]{Java}{
  morekeywords={conforms, to, tags, for, tag, with}
}
\lstdefinelanguage{cd}[]{Java}{
  morekeywords={classdiagram, String, Date, association}
}
\lstdefinelanguage{gui}[]{Java}{
  morekeywords={webpage, container, card, head, body, datatable, rows, column, button, row}
}

\usepackage{xspace}
\usepackage{ifdraft}
\usepackage{csquotes}

\usepackage{prettyref}
\newrefformat{chap}{Chapter~\ref{#1}}
\newrefformat{sec}{Section~\ref{#1}}
\newrefformat{subsec}{Section~\ref{#1}}
\newrefformat{eq}{Equation~(\ref{#1})}
\newrefformat{fig}{Figure~\ref{#1}}
\newrefformat{ex}{Example~\ref{#1}}
\newrefformat{list}{List~\ref{#1}}
\newrefformat{tab}{Table~\ref{#1}}
\newrefformat{enum}{Case~(\ref{#1}.)}
\newrefformat{def}{Definition~\ref{#1}}
\newrefformat{le}{Lemma~\ref{#1}}
\newrefformat{th}{Theorem~\ref{#1}}

\makeatletter
\newcommand*{\etc}{%
  \@ifnextchar{.}%
  {\textit{etc}}%
  {\textit{etc.}\@\xspace}%
}
\makeatother

\definecolor{se-green}{RGB}{0,128,0}
\definecolor{se-blue} {RGB}{0,0,204}

\newcounter{requirement}[section]

\usepackage{tabularx}
\usepackage{booktabs}
\usepackage{makecell}

\newcolumntype{C}{>{\centering\arraybackslash}X}
\newcolumntype{D}[1]{>{\centering\let\newline\\\arraybackslash\hspace{0pt}}m{#1}}
\newcolumntype{M}[1]{>{\centering\arraybackslash}m{#1}}
\usepackage{multirow}
\usepackage{multicol}
\usepackage{threeparttable}

\usepackage{subfig}

\AtBeginDocument{%

}

\newacronym{AST}{AST}{Abstract Syntax Tree}

\newacronym{BPMN}{BPMN}{Business Process Model and Notation}

\newacronym{BPD}{BPD}{Business Process Diagram}

\newacronym{CTL}{CTL}{Computation Tree Logic}

\newacronym{CD}{CD}{Class Diagram}

\newacronym{CoCo}{CoCo}{Context Condition}

\newacronym{CPS}{CPS}{Cyber-Physical System}

\newacronym{DTO}{DTO}{Data Transfer Object}

\newacronym{DT}{DT}{Digital Twin}

\newacronym{DSL}{DSL}{Domain-Specific Language}
\def\dsl{\gls{DSL}\xspace}
\def\dsls{\glspl{DSL}\xspace}

\newacronym{EIS}{EIS}{Enterprise Information System}

\newacronym{GUI}{GUI}{Graphical User Interface}

\def\guis{\glspl{GUI}\xspace}

\newacronym{GPL}{GPL}{General Purpose Language}
\def\gpl{\gls{GPL}\xspace}
\def\gpls{\glspl{GPL}\xspace}

\newacronym{IoT}{IoT}{Internet of Things}

\newacronym[longplural={Lines of Code}]{LOC}{LOC}{Line of Code}

\newacronym{MDSE}{MDSE}{Model-Driven Software Engineering}

\newacronym{MDE}{MDE}{Model-Driven Engineering}
\def\mde{\gls{MDE}\xspace}

\newacronym{MOF}{MOF}{Meta Object Facility}
\def\mof{\gls{MOF}\xspace}

\newacronym{OCL}{OCL}{Object Constraint Language}

\newacronym{OD}{OD}{Object Diagram}

\newacronym{OM}{OM}{Object Model}

\newacronym{PIM}{PIM}{Platform Independent Model}
\def\pim{\gls{PIM}\xspace}
\def\pims{\glspl{PIM}\xspace}

\newacronym{PSM}{PSM}{Platform Specific Model}
\def\psm{\gls{PSM}\xspace}
\def\psms{\glspl{PSM}\xspace}

\newacronym{PAIS}{PAIS}{Process-Aware Information System}

\newacronym{RTE}{RTE}{Runtime Environment}

\newacronym{UML}{UML}{Unified Modeling Language}
\def\uml{\gls{UML}\xspace}
\glsunset{UML}

\newacronym{UML/P}{UML/P}{\glspl{UML}/P}

\newacronym{UI}{UI}{User Interface}
\def\ui{\gls{UI}\xspace}

\newacronym{WCAG}{WCAG}{Web Content Accessibility Guidelines}
\def\wcag{\gls{WCAG}\xspace}

\usepackage{ifdraft}
\usepackage{ifthen}

\newboolean{debug} 
\setboolean{debug}{false} 

\ifthenelse{\boolean{debug}}{%
  \usepackage{showframe} 
  \usepackage{hyperref}  
  \pagestyle{plain}      
  \newcommand{\xynote}[2]{\todo[inline]{#1: #2}}
  
}{%
  \usepackage[disable]{todonotes}        
  \hypersetup{hidelinks} 
  \pagestyle{empty}
  \newcommand{\xynote}[2]{}
  
}

\AtBeginDocument{%

}

\AtBeginDocument{%
  \providecommand\BibTeX{{%
    \normalfont B\kern-0.5em{\scshape i\kern-0.25em b}\kern-0.8em\TeX}}}

\begin{document}
\title{Addressing Visual Impairments with Model-Driven Engineering: A Systematic Literature Review}


\author{Judith Michael}
\orcid{0000-0002-4999-2544}
\email{judith.michael@ur.de}
\affiliation{%
  \institution{University of Regensburg}
  \city{Regensburg}
  \country{Germany}
}
\author{Lukas Netz}
\orcid{0000-0003-2013-2919}
\email{netz@se-rwth.de}
\author{Bernhard Rumpe}
\orcid{0000-0002-2147-1966}
\email{rumpe@se-rwth.de}
\affiliation{%
  \institution{Software Engineering, RWTH Aachen University}
  \city{Aachen}
  \country{Germany}
}

\author{Ingo Mueller}
\orcid{0000-0003-2240-712X}
\email{ingo.mueller@monashhealth.org}
\affiliation{%
  \institution{Monash Health}
  \city{Melbourne}
  \country{Australia}
}

\author{John Grundy}
\orcid{0000-0003-4928-7076}
\email{john.grundy@monash.edu}
\author{Shavindra Wickramathilaka}
\orcid{0000-0002-4732-2264}
\email{shavindra.wickramathilaka@monash.edu}
\affiliation{%
  \institution{Monash University}
  \city{Melbourne}
  \country{Australia}
}

\author{Hourieh Khalajzadeh}
\orcid{0000-0001-9958-0102}
\affiliation{%
  \institution{Deakin University}
  \city{Melbourne}
  \country{Australia}
}
\email{hkhalajzadeh@deakin.edu.au}

\renewcommand{\shortauthors}{J. Michael, L. Netz, B. Rumpe, I. Müller, J. Grundy, S. Wickramathilaka, H. Khalajzadeh}

\begin{abstract}
Software applications often pose barriers for users with accessibility needs, e.g., visual impairments. Model-driven engineering (MDE), with its systematic nature of code derivation,  offers systematic methods to integrate accessibility concerns into software development while reducing manual effort. This paper presents a systematic literature review on how MDE addresses accessibility for vision impairments. From 447 initially identified papers, 30 primary studies met the inclusion criteria. About two-thirds reference the Web Content Accessibility Guidelines (WCAG), yet their project-specific adaptions and end-user validations hinder wider adoption in MDE. The analyzed studies model user interface structures, interaction and navigation, user capabilities, requirements, and context information. However, only few specify concrete modeling techniques on how to incorporate accessibility needs or demonstrate fully functional systems. Insufficient details on MDE methods, i.e., transformation rules or code templates, hinder the reuse, generalizability, and reproducibility. Furthermore, limited involvement of affected users and limited developer expertise in accessibility contribute to weak empirical validation. Overall, the findings indicate that current MDE research insufficiently supports vision-related accessibility. Our paper concludes with a research agenda outlining how support for vision impairments can be more effectively embedded in MDE processes.
\end{abstract}



\maketitle

\section{Introduction}
\label{sec:intro}

Developers often overlook that modern software applications are not only technical solutions but are embedded in a larger socio-technical systems including diverse end users with heterogeneous needs. 
In this article, we focus on a particular aspect of accessibility needs, visual impairments. 
To give an example from practice: When selecting the menu for lunch or reading a bus schedule, users with low vision or blind users have specific requirements for the technical applications used, e.g., showing larger text sizes and having more contrast for people with low vision or enabling the reading of texts with screen readers for blind users.  
This requires developers of such systems to consider these needs within the engineering of socio-technical systems; not only due to their strive for inclusion but also to fulfill legal requirements for providing accessible systems, i.e., the European accessibility act~\cite{EUAA25}. 

Ensuring the accessibility of \guis is particularly important for people with visual impairments.
GUIs are the main medium through which users interact with software applications, and their accessibility has a direct impact on the usability and comprehensibility of digital content. For people with visual impairments, inclusive GUIs facilitate equal access to information, promote independence, and social inclusion.
Descriptions on how to cover needs for visual impairments are prominently featured in accessibility guidelines such as \wcag~\cite{WCAG2.2}. 
Accessibility as a non-functional requirement, however, is challenging to realize in a software system because its effects are widespread across the entire system~\cite{sommerville2018}. 


\mde claims to have various advantages useful for addressing accessibility in applications~\cite{VSB+13}. This includes better reusability and portability, increasing development speed, increasing software quality due to uniform representations in the code, and being better maintainable when it comes to technological changes. Cross-cutting aspects, including UX and accessibility, can be changed in just one place -- the models. When generating code, this change is then reflected consistently in many different places in the code. 
Many \mde methods have been adopted in practice ~\cite{Brambilla17,Acerbis07}, e.g., for design automation~\cite{GKM+25}, code generation~\cite{Vogel14,Koziolek20,BGK+24}, test automation~\cite{Rutherford03,Baker05}, automated verification and validation~\cite{Colantoni21,Mohagheghi13,KMP+21}, or simulation~\cite{Hutchinson14,Eramo24}.
\mde enhances the software engineering process by increasing automation through the use of \gpls or \dsls, as well as automated code generation, transformations, or interpretation. 
On top of \mde~\cite{DiRuscio22}, low-code development platforms have moved \mde  from research to practice, providing methods for supporting additional stakeholder groups~\cite{Alfonso24,DHM+22}.

Even though \mde seems to be a promising approach for providing more accessible applications, how is it used to improve applications for visually impaired users?
Even though standards and suggestions for developers have existed for many years, e.g., \wcag~\cite{WCAG2.2}, User Agent Accessiblity Guidelines~\cite{UAAG}, or \emph{EN 301 549}~\cite{EN301549}, their application in practice seems to be still limited. 
Widely adopted GUI frameworks provide suggestions on how to tackle it, but there exists no automated support to cover accessibility. Existing techniques lack enough support for agile software development including iterations to improve end user reported issues like accessibility~\cite{79}. 
In this paper, we report the results of a Systematic Literature Review (SLR) 
to identify the current state-of-the-art of addressing accessibility needs with \mde \textit{based on the concrete example of vision impairments}.
The main contributions of this work include:
\begin{itemize}
    \item a detailed examination of 30 highly relevant studies shedding light on the publication venue, timing, output, and nature of reported approaches,
    \item an analysis of which visual impairments are addressed with model-driven approaches;
    \item analysis of  how \mde approaches for the design and implementation of systems covering visual impairment needs work;
    \item  analysis of how the developed approaches have been evaluated;
    \item summarisation of the approaches' reported strengths, limitations, gaps and challenges;
    \item the main limitations of the reported studies; and
    \item a research roadmap covering different aspects that we recommend be investigated in the future.
\end{itemize}

The article is structured as follows: \autoref{sec:background} provides the relevant background and related literature reviews. \autoref{sec:method} describes our research method. 
\autoref{sec:results} presents the main results of the analysis based on the posed research questions and highlights the main findings. 
\autoref{sec:discussion} lists the identified limitations and the research roadmap, while
\autoref{sec:threats} discusses the main threats to validity. 
\autoref{sec:conclusions} concludes. 
\section{Background and Related Work}
\label{sec:background}


In this section, we introduce the concepts of software accessibility and model-driven engineering, set the scope of this study to visual impairments, and discuss related literature reviews.

\subsection{Software Accessibility and Visual Impairments}
\label{sec:accessibility}

\emph{Software accessibility} refers to the design of software systems that ensures barrier-free use by all, including people with disabilities. The concept of accessibility has been explored in a software engineering context as early as the mid 1990s (See Laux et al. \cite{Faux+96} for an example). These efforts have evolved into several technical standards including the \emph{\wcag}~\cite{WCAG2.2}, \emph{User Agent Accessibility Guidelines (UAAG)}~\cite{UAAG}, \emph{Authoring Tool Accessibility Guidelines (ATAG)}~\cite{ATAG}, or \emph{EN 301 549}~\cite{EN301549} to name a few. Furthermore, software accessibility has gained significance in the IT industry over the last decade, particularly as it has been legislated in several jurisdictions as a mandatory requirement for public sector Web sites and mobile applications such as the US (21st Century Communications and Video Accessibility Act (CVAA)~\cite{CVAA21}), EU (Directive 2016/2102~\cite{EUDirective16}), and Australia (Web Accessibility National Transition Strategy~\cite{Australia10}).

Although accessibility encompasses ``[...] a wide range of disabilities, including visual, auditory, physical, speech, cognitive, language, learning, and neurological disabilities.'' \cite{WCAG2.2}, this research focused on \emph{visual impairments} only. We have chosen this narrow scope because we are \underline{not} interested in studying the coverage (completeness of implementation) of accessibility standards such as WCAG, but rather aim to drill down on a) the maturity of available tools and methods for the implementation of a concrete real-world accessibility need, as well as b) the benefits to the end users. Moreover, we decided to focus on visual impairments because they do not only affect people with disabilities but are experienced by a larger segment of the general population such as the elderly. 


Common visual impairments as reported by the Australian Institute of Health and Welfare (AIHW)~\cite{AIHW21} are:
\begin{itemize}
    \item astigmatism (imperfection in the curvature of the eye)
    \item blindness (complete and partial)
    \item cataract (cloudy area in the lens)
    \item color blindness
    \item hyperopia (long-sightedness)
    \item macular degeneration (eye disease that affects central vision)
    \item myopia (short-sightedness)
    \item presbyopia (loss of focusing ability with age).
\end{itemize}

All listed impairments except color-blindness and (complete and partial) blindness are typically referred to in the SE literature under the umbrella term of visual, or vision, impairments. The accessibility needs related to visual impairments are complex. Firstly, the above list is by no means exhaustive. Secondly, the severity of visual impairments differs from person to person and, thirdly, often changes with time or situation. It is therefore not trivial to design software systems that address visual impairments. Standards like WCAG provide guidance, however considerable effort on the part of the developers is still required to implement accessible software systems. A set of easy-to-use tools and techniques may simplify the situation and remove barriers for developers to implement accessible software systems. Model-driven engineering approaches have the potential to do exactly this. 


\subsection{Model-driven Engineering}
\label{sec:mde}

\mde is a software development methodology that enables the generation of software systems based on the creation and transformation of models. Information about a given problem (domain) is typically provided with a visual or textual \emph{Domain Model}, ``which describes the various entities, their attributes, roles, and relationships, plus the constraints and interactions that describe and grant the integrity of the model elements [...]'' \cite{brambilla+17}. Hence, \mde is centered on modeling the concepts of the respective problem domain, instead of the details of the programmatic implementation of a solution.

The other core \mde concept --- \emph{transformations} --- are well-defined mappings between source and target models. Transformations enable the step-wise refinement of an abstract model (the domain model) to a concrete executable model, or in other words, source code. Executable models typically come in one of two forms: a) code generation with an existing programming language (also known as model-to-text transformation), or b) virtual machine-based interpretation of a low-level representation of a model. \footnote{A detailed discussion of \mde principles is beyond the scope of this study. Please refer to Brambilla et al. \cite{brambilla+17} for a comprehensive introduction.}

\mde claims to offer several key advantages~\cite{VSB+13} including
\begin{itemize}
    \item Openness. As domain models represent concepts of a problem domain, even domain experts with limited to no programming skills can use MDE tools to create software systems.
    \item Separation of concerns. Different models can be used to capture different aspects of a problem domain separately to break down complexity.
    \item Automation. Once models and transformations are defined, software can be generated with limited to no need for manual intervention.
\end{itemize}

If it is possible to capture accessibility needs of visually impaired users with a dedicated domain model (cf. Digital Human Twins~\cite{Bra25}), and to define transformation rules to map accessibility needs into models of a concrete problem domain until executable, it may be possible to streamline and automate the implementation of accessibility needs for users with visual impairments. 
While there exist approaches that consider MDE and accessibility for seniors~\cite{wickramathilaka2025adaptive}, for accessible web pages~\cite{ordonez2024uml}, automated test kits for accessibility in Android apps~\cite{garciaearly2023}, or MDE and accessibility in general~\cite{BKS+25, Matos25}, the specific area of MDE and accessibility for visual impairments is less researched.


\subsection{Related Literature Reviews}
\label{sec:relatedreviews}


The concept of accessibility has attracted interest in SE since the mid 1990s. The roots of MDE can be traced back even further to the emergence of computer aided software engineering (CASE) tools in the 1980s. Consequently, a plethora of primary studies on either topic has been published over the years. The objective of this research is to identify those studies that address a combination of both, or more specifically, the design and implementation of software systems that address the needs of people with visual impairments using MDE methods. This article reports on the systematic and rigorous identification and assessment of relevant primary studies.


As a first step, we identified the need for our study by reviewing existing secondary studies that already address our area of interest. At the time of writing, only one highly relevant study was available. Ordo\~{n}ez et al. \cite{ordonez2022} offer a well-presented review of 37 primary studies based on the following research questions:

\begin{enumerate}
    \item What are the motivations for using MDD for producing accessible software?
    \item Which method is proposed for the model-driven development of accessible software?
    \item Which tools and languages provide technological support for the integration of accessibility in the context of model-driven development?
    \item What are the advantages offered by the model-driven development of accessible software?
    \item What are the disadvantages in the context of model-driven development of accessible software?'' \cite{ordonez2022}
\end{enumerate}

While Ordo\~{n}ez et al. focus on the intersection of MDE and accessibility, they present a generic mapping of reported approaches, rather than a detailed analysis of the methods and tools used to model and generate accessible software systems. Despite being a valuable contribution, it is hard to assess the applicability of the reviewed approaches to address concrete real-world accessibility needs. In contrast our SLR will go more into detail and provide:

\begin{itemize}
    \item An in-depth evaluation of the capabilities of approaches to capture accessibility (visual impairment) needs,  
    \item A detailed analysis of the evaluations of reported approaches and benefits to end users, and 
    \item A detailed assessment of the maturity and reproducibility of reported approaches.
\end{itemize}

Seven additional partially relevant secondary studies were identified. De Oliveira and Filgueiras~\cite{OF18} prepared a systematic review of development tools that support the creation of mobile applications accessible to visually impaired users. The focus of this study is on practical tools rather than methods with MDE being out of scope. Rokis and Kirikova \cite{rokis2022} as well as Bucaioni et al. \cite{bucaioni2022} review challenges of \emph{Low Code} software development without considering accessibility. 
The following studies are relevant regarding accessibility and low vision but have no focus on MDE:
Di Gregorio et al. \cite{digregorio2022} review practical aspects of the creation of accessible mobile apps. Paiva et al.~\cite{Paiva+21} provide a review of the coverage of accessibility needs in software engineering processes. 
Teixeira et al.~\cite{Teixeira24} analyze how accessibility requirements are considered in the development process of information systems. 
De Souza et al.~\cite{Souza2024} analyze the use of intelligent environments for visually impaired people. 

\section{Research Method}
\label{sec:method}

\begin{figure}[h]
    \centering
    \includegraphics[width=0.7\linewidth]{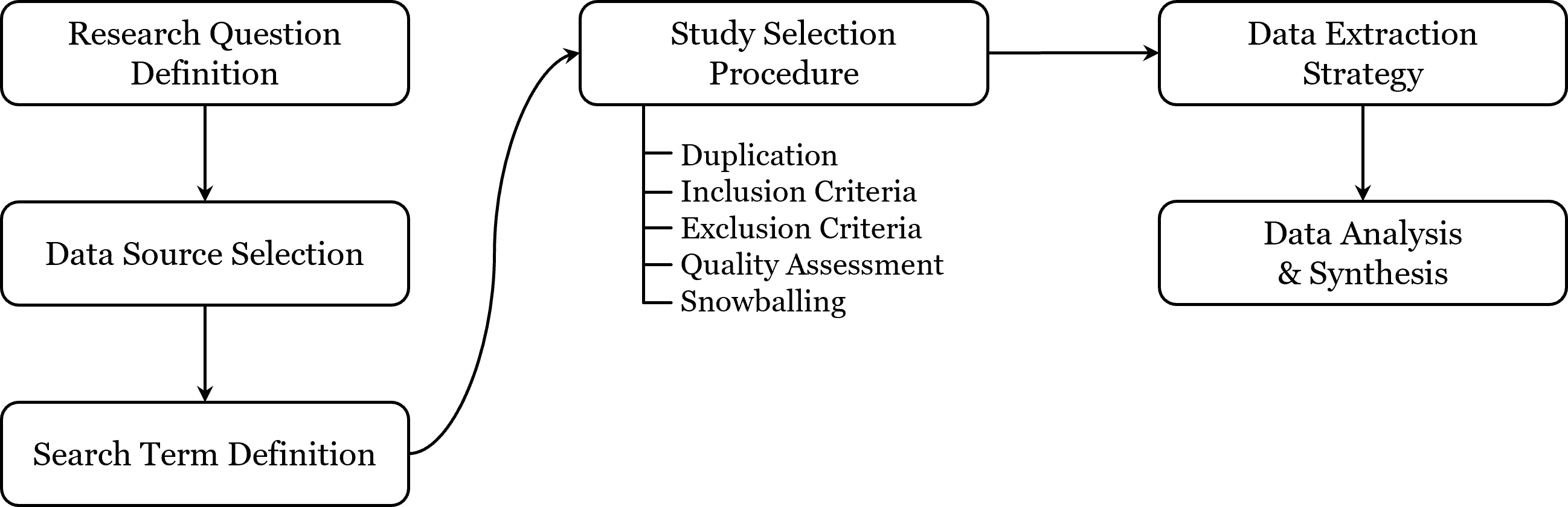}
    \caption{Overview of our research process design and its key components, namely \textit{Research Questions} (cf. \autoref{sec:RQ}), \textit{Data Sources} (cf.\autoref{sec:datasources}), \textit{Search Term} (cf. \autoref{sec:searchterm}), \textit{Study Selection} (cf. \autoref{sec:selectionprocedure}), \textit{Data Extraction} (cf. \autoref{sec:dataExtraction}) and \textit{Data Analysis} (cf. \autoref{sec:analysis-synthesis})}
    \label{fig:research-process-overview}
\end{figure}

Our SLR followed the widely accepted guidelines for the preparation of systematic SE literature reviews by Kitchenham et al. \cite{kitchenham2023}, as well as the Wohlin \cite{wohlin2014} guidelines for a systematic snowballing strategy. An overview of the design of our SLR process is depicted in Figure \ref{fig:research-process-overview}. The remainder of this section details the key elements of this process.

\subsection{Research Questions}
\label{sec:RQ}
The key research questions (RQ) addressed by this study are:

\textbf{RQ1.  Which trends exist in terms of publication date, venue, impact and nature of reported approaches?} -- This RQ explored the visibility and impact of the selected studies in the MDE community. For this reason, we analyzed publication frequency, types and venues, looked into author affiliations and citation counts, and assessed the main  contribution type of each of the selected studies.

\textbf{RQ2. What visual impairments are addressed with model-driven approaches?} -- The aim of this RQ was to understand the depth and breadth with which human aspects of visual impairment needs were addressed in the selected studies, in the context of varying user needs and their complexities for different visual impairments.

\textbf{RQ3. How do model-driven approaches for the design and implementation of visual impairment needs work?} -- This RQ examined the technical details of the selected studies. We analyzed the reported approaches in depth and present our findings using the stages of a software development life cycle. 


\textbf{RQ4. How are these approaches evaluated?} -- This RQ reviewed the evaluation methods documented in the selected studies, with a specific focus on user acceptance testing with persons with visual impairments, or in other words, how well visual impairment needs were addressed by the selected studies. 

\textbf{RQ5. What strengths, limitations, gaps and challenges are reported?} -- With this RQ we identified the limitations of the selected studies as well as open challenges and future work in the domain of MDE for the design and implementation of visual impairment needs.


\subsection{Data Sources}
\label{sec:datasources}
We decided to search for literature in well-established scientific online databases. We chose to use:

\begin{itemize}
    \item ACM Digital Library, \url{https://dl.acm.org/}
    \item IEEE Xplore, \url{https://ieeexplore.ieee.org/}
    \item ScienceDirect, \url{https://www.sciencedirect.com/}
    \item Scopus, \url{https://www.scopus.com/}.
\end{itemize}

These selected databases provide access to rigorously peer-reviewed publications --- an important quality aspect reflected in our exclusion criteria. ACM and IEEE were included because they are primary venues for the dissemination of SE / MDE literature. Scopus and ScienceDirect were added to widen the search to relevant studies in other disciplines. Note, our original design included Springer (\url{https://link.springer.com/}) as well. However, a large number of irrelevant hits combined with the export limit of 1000 records made the use of Springer ineffective. Instead, we opted for a snowballing procedure to ensure the detection of further studies that did not produce hits in the selected databases (cf. Sec. \ref{sec:snowballing}).


\subsection{Search Term}
\label{sec:searchterm}
The search was performed based on below search string that contained several generic elements to ensure both aspects of our study --- visual impairments and MDE --- were present in all candidate studies:

\begin{center}
(accessibility OR blind* OR "low vision" OR "vision impair*" OR "visual* impair*") \\ AND (MDE OR "model-driven")
\end{center}

Note that the omission of the conditions listed in Sec~\ref{sec:accessibility} was due to the fact that they are typically referred to in the SE literature under the umbrella term of `visual impairment' or related synonyms. The keyword `accessibility' was added to ensure the detection of candidate studies that do not mention visual impairments in the searched parts.

Logical operators ensured that the two different aspects of our study (accessibility and MDE) were both present (\emph{AND}), and at least one alternative spelling of a keyword or synonym was reflected (\emph{OR}) in all search hits. Parentheses \emph{()} allowed logical groupings, apostrophes \emph{""} marked atomic terms, while wildcards \emph{*} were used to account for alternative spellings.

The search query was tested and refined with a series of pilot runs. The result of each run and subsequent refinements were discussed in several team meetings with all authors present. The final database-specific search parameters are presented in Table \ref{tab:databases} in Appendix \ref{sec:rm-appendix}. The search was carried out by one author and candidate studies were documented in a spreadsheet.



\subsection{Study Selection Procedure}
\label{sec:selectionprocedure}
The selection procedure encompassed multiple screenings to remove duplicates and to apply inclusion and exclusion criteria. These screenings were performed by one author who documented all rejected studies incl the reason for their rejection in a spreadsheet. Five percent of rejected studies were randomly assigned to other authors for cross-checking. 


\subsubsection{Deduplication}
The removal of all duplicates of candidate studies after the online search.

\subsubsection{Inclusion Criteria}
Note, our SLR did \underline{not} place restrictions on the publication date to ensure all relevant previous and current contributions were detected. Likewise, we kept our study open to all scientific venues, including journals, conferences and workshops. Candidate studies were included if they satisfied all of the following conditions:

\begin{itemize}
    \item Study reported a model-driven or related model-based approach
    \item Study focused on accessibility needs
    \item Study presented a model-driven method to address a visual impairment need. 
\end{itemize}

Model-based approaches were considered \emph{related} if they cover or contribute to model-driven techniques or methods such as the modeling of accessibility needs (e.g. with a domain-specific language and/or visual notation).

The inclusion criteria were applied by checking study titles, abstracts and author keywords for (a) direct keyword matches, and (b) semantic relevance. Semantic relevance was defined to exist if a study related to the key aspects of this SLR --- visual impairments/accessibility and MDE. If no match was detected, the study was dropped.

\subsubsection{Exclusion Criteria}
The following types of studies were excluded
\begin{itemize}
    \item Non-primary studies 
    \item Non-English language studies
    \item Non-peer reviewed or non-verifiably peer-reviewed studies 
    \item Studies superseded by newer versions
    \item Short papers of 4 or less pages incl. references.
\end{itemize}

Note, short papers were excluded because they do not provide sufficient detail due to space limitations.

\subsubsection{Quality Assessment}
\label{sec:initial_review}

The quality of each candidate study was evaluated based on below questions and the scoring system presented in Table \ref{tab:qascoringsystem}. The results of our quality assessment are presented in the appendix in Table \ref{tab:QA}. Although none of the studies reached the maximum score (indicating very high quality), we decided not to remove studies based on their score due to the small number of identified candidate studies and to reduce publication bias.


\begin{itemize}
    \item[QA1] Was the study clearly motivated by a real-world visual impairment need?
    \item[QA2] Was the presented approach clearly defined?
    \item[QA3] Was sufficient detail provided to enable reproduction and replication?
    \item[QA4] Was an accessibility evaluation with visually impaired users presented?
\end{itemize}

\begin{table}[h]
\caption{\label{tab:qascoringsystem} Quality Assessment (QA) scoring system}
\footnotesize
\resizebox{\textwidth}{!}{%
\begin{tabular}{|l|p{4cm}|p{4cm}|p{4cm}|}
 \hline
    Criterion & \multicolumn{3}{c}{Score} \\
     & Y (yes) -- 1.0 & P (partial) -- 0.5 & N (no) -- 0.0 \\\hline
    QA1 & concrete visual impairment needs stated & generic motivation, visual impairments mentioned & motivation without visual impairments \\\hline
    QA2 & solution concept clearly and formally defined & solution concept clearly but informally defined & solution concept is unclear or no concept given \\\hline
    QA3 & detail sufficient and clear, reproduction possible & detail somewhat sufficient and clear, reproduction partially possible with assumptions & detail either insufficient or unclear, reproduction not possible \\\hline
    QA4 & acceptance testing with visually impaired users reported & acceptance testing with partially / fully sighted users reported & weaker forms of evaluation or no evaluation presented \\\hline
\end{tabular}}
\end{table}


The quality assessment of the candidate studies was distributed evenly among all authors. We met several times to discuss individual findings, identify discrepancies, and cross-check our reviews. For the cross-checks, each author was randomly assigned three studies that were originally reviewed by another team member. The results of these second reviews were discussed and agreed in two meetings.

\subsubsection{Snowballing Procedure}
\label{sec:snowballing}
All selected studies underwent forward and backward snowballing as per Wohlin \cite{wohlin2014}:
\begin{itemize}
    \item\textbf{backward snowballing} ``means using the reference list [of a candidate study] to identify new study to include''
    \item\textbf{forward snowballing} ``refers to identifying new studies based on those studies citing the study being examined''.
\end{itemize}

Forward snowballing was carried out with Google Scholar \url{https://scholar.google.com.au/} which provides access to forward references of a study via the \emph{`Cited by'} link. Google Scholar was chosen because it surpasses our selected databases in providing the most up-to-date list of `Cited by' references based on information retrieved from a wide range of online databases and other online sources. 

The set of snowballed candidate studies was divided up. Identified studies were discussed and agreed upon in meetings with all authors present. A pre-selection of candidate studies that `simulated' the database search by title, abstract and keywords was carried out to reduce the number of candidates to be evaluated. Snowballed studies were treated like all other candidate studies and underwent the same selection procedure. Snowballing was continued until no further candidates were found.


\subsection{Search Results and Study Selection}
\label{sec:literature_search}



Our initial literature search was performed in February 2023 and repeated in January 2024 and we performed an additional forward snowballing in October 2025.
The search results per database and the total number of 207 identified candidate studies are shown in Table~\ref{tab:searchAndSelection}. The number of candidates with Scopus was dis-proportionally higher than with other databases. The likely reason for this is that Scopus provides access to research literature from a diverse range of disciplines where some of our search keywords may have a different meaning. However, the extra number of candidates did not pose a problem for our SLR, except for requiring additional time during the selection process. An additional 240 candidates were identified during two iterations of snowballing (iteration I: 136, iteration II: 104). Table \ref{tab:searchAndSelection} also provides an overview of our study selection results. Columns 2--4 show the number of removed candidate studies for each step of the selection process, while column 5 shows the number of selected studies. After completing study selection, a total of 30 primary studies were selected for our SLR (cf. Table \ref{tab:selectedStudies} in Appendix \ref{sec:selected-study-appendix}).

\begin{table}[h]
\caption{\label{tab:searchAndSelection} Overview of search / snowballing candidates and study selection results}
\footnotesize
\begin{tabular}{|lr|r|r|r|r|}
 \hline
    \multicolumn{2}{|l|}{Total number of candidates} & \multicolumn{3}{l}{Number of removed candidate studies} \vline & \multicolumn{1}{l}{Number of} \vline \\
    & & Deduplication & Inclusion Criteria & Exclusion Criteria & selected studies \\\hline\hline
    Database search & & & & & \\ 
    \small ACM DL & \small 38 & & & & \\
    \small IEEE Xplore & \small 32 & & & & \\
    \small ScienceDirect & \small 15 & & & & \\
    \small Scopus & \small 122 & & & & \\
    & 207 & 56 & 102 & 30 & \bf{19} \\\hline
    Snowballing, iteration I & 136 & 61 & 41 & 23 & \bf{11}\\\hline
    Snowballing, iteration II & 104 & 82 & 15 & 7 & \bf{0}\\\hline
    \multicolumn{6}{r}{\bf{total number of selected studies: 30}}
\end{tabular}
\end{table}

\subsection{Data Extraction Process}
\label{sec:dataExtraction}
A comprehensive set of data points was defined for extracting the necessary information to answer our RQs (cf. Sec.~\ref{sec:RQ}). These data points were grouped into the four categories: \emph{Metadata}, \emph{Accessibility}, \emph{Approach and results}, and \emph{Evaluation, limitations and open challenges} as shown in Table \ref{tab:dataPoints} in Appendix \ref{sec:rm-appendix}. Data extraction was conducted by all five authors and collected in a shared spreadsheet. The extraction process started with three joint sessions to establish a shared understanding and vocabulary, share data extraction best practices, and resolve open questions and anomalies. Each author was randomly assigned three studies upon completion for cross-checking.


\subsection{Data Analysis and Synthesis}
\label{sec:analysis-synthesis}
The extracted data were analyzed and synthesized using quantitative and qualitative methods. For example, the data in the metadata category such as place and year of publication were quantitatively summed up. In contrast, thematic analyzes were conducted on the data collected using data points DP4.5 and DP4.6. All selected studies were first analyzed individually to identify their key features. A horizontal analysis across all selected studies was then carried out to identify patterns and trends, and to group and classify individual studies. Data analysis and synthesis were carried out in groups of at least two authors. Results and findings were discussed and agreed on in multiple team meetings.

\section{Results}
\label{sec:results}





\subsection{RQ1 -- What trends exist in terms of timing, output, impact and nature of reported approaches?}
\label{subsec:RQ1}

Our first RQ explores the visibility and impact of the selected studies in the MDE community. For this reason, we analyzed publication frequency, types and venues, looked into author affiliations and citation counts, and also assessed the problem and contribution types of each of the selected studies.

\subsubsection{Timing and output} 
\label{sec:timing}
\begin{figure}[h!]
    \centering
    \subfloat[Number of publications per year\label{fig:pubyears}]{
        \includegraphics[width=0.45\textwidth]{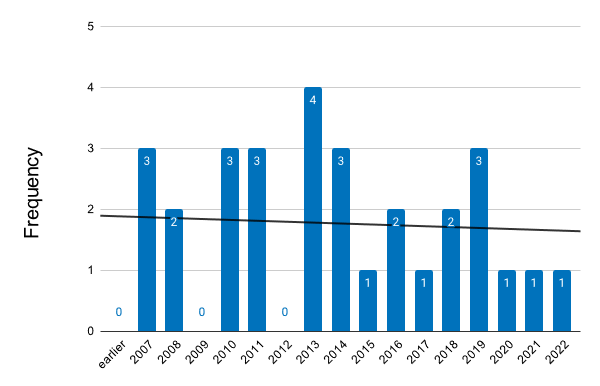}
    }
    \hfill
    \subfloat[Number of citations (histogram + normal distribution)\label{fig:histogram}]{
        \includegraphics[width=0.45\textwidth]{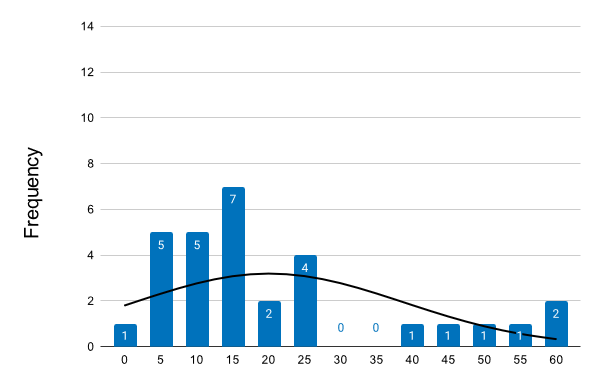}
    }
    \caption{Comparison of publications and citations over the years}
    \label{fig:comparison}
\end{figure}

As shown in Figure \ref{fig:pubyears}, our 30 selected studies started appearing from 2007 onward with an overall low frequency that is slowly trending downward after a more active period between 2007 to 2014. On this basis, three observations can be made:
\begin{enumerate}
    \item \emph{Delayed interest.} The MDE community started covering the topic of visual impairments significantly later than the SE domain in general (2007 vs mid 1990s) (cf. Section \ref{sec:accessibility}).
    \item \emph{Low output.} MDE approaches that address visual impairments are a niche topic with a peak of only 4 publications in 2013.
    \item \emph{No clear trend.} An increasing accumulation of knowledge (publications) over the years cannot be observed. On the contrary, the momentum has dropped off since 2019.
\end{enumerate}

\begin{tcolorbox}[colback=white, colframe=white, boxrule=0mm, sharp corners]
    \textit{Research into MDE approaches that address visual impairment needs started late, trends downward, and has yielded a low output to date.}
\end{tcolorbox}

\subsubsection{Impact} 
\label{sec:impact}

To determine the impact in the MDE community, we examined the publication venues, author affiliations, and citation counts of our selected studies. 
While publication venues show a large spread (cf. Table \ref{tbl:pub_venues} in Appendix \ref{sec:results-appendix}), surprisingly none of the 30 selected studies was published at a high-ranked MDE venue such as MODELS, ASE, SLE, GPCE or SoSyM. Instead a trend towards Human-Computer Interaction (HCI)-related journals and conferences can be observed, e.g. the Journal on Universal Access in the Information Society (3 studies), ACM International Conference on Design of Communication (3 studies), Conference on Innovations in E-learning, Instruction Technology, Assessment and Engineering Education (2 studies), and the International Conference on Universal Access in Human-Computer Interactions (2 studies).

Our 30 selected studies were created by a small community of authors who frequently collaborate with each other. Table \ref{tbl:affiliations} in Appendix \ref{sec:results-appendix} maps all 23 author teams against their published studies and depicts which of the studies were co-authored by multiple teams. The table also shows that only nine author teams have published more than one study on the topic of this SLR -- often in collaboration with other teams. 
The dissemination of information beyond this community of authors is to date limited, as the low citation counts indicate. We used Google Scholar to determine the number of citations per selected study. Figure \ref{fig:histogram} depicts a histogram of the citation frequency. As can be seen, the largest portion of studies (7 studies) was cited between 11-15 times. Only a small number of studies (6) were cited more than 30 times as of 1 July 2025. 
This may be due to the specific focus of our selected studies, but can also be interpreted as a trend towards low impact in the MDE community.


\begin{tcolorbox}[colback=white, colframe=white, boxrule=0mm, sharp corners]
    \textit{The impact in the MDE community to date is low. None of the selected studies appeared in a high ranked MDE venue. Generally low citation counts indicate that most of the studies did not create interest beyond a small community of collaborating author.}
\end{tcolorbox}

\subsubsection{Problem Approach, Contribution and Study Type} 

With respect to the nature of the 30 selected studies, we analyzed information about their problem approach, the type of contribution and study type. 
Regarding the overall problem approach, 29 out of 30 selected studies describe research under lab conditions, which typically entails a lack of motivation with a concrete real-world need, and/or a lack of evaluation with vision impaired end users. This reflects the often abstract and conceptual nature of the majority of the selected studies. Only study S6 presents and evaluates work under realistic real-world conditions.

A majority of 20 studies presents a new method or technique, that is in just under 50 per cent of cases accompanied by a prototype demonstration or a tool description (cf. Table \ref{tbl:contribution_types}). In contrast, only 5 out of 30 studies provide an evaluation with end users -- vision impaired or not. Hence a focus on technical aspects can be noticed.



\begin{table}[h]
\centering
\begin{minipage}{0.45\linewidth}
\centering
\caption{Overview of contribution types}
\label{tbl:contribution_types}
\footnotesize
\begin{tabular}{|l|r|}
\hline
Contribution type & Number of\\
 & studies\\\hline\hline
Method / technique & 20 \\\hline
Tool & 3 \\\hline
Prototype & 9 \\\hline
Evaluation & 5 \\\hline
Replication study & 0 \\
\hline
\end{tabular}
\end{minipage}
\hspace{0.05\linewidth}
\begin{minipage}{0.45\linewidth}
\centering
\caption{Overview of Study types}
\label{tbl:pubtypes}
\footnotesize
\begin{tabular}{|l|r|}
\hline
Study type & Number of\\
\hline\hline
Journal papers & 12 \\\hline
Regular conference paper & 8 \\\hline
Short conference paper & 8 \\\hline
Extended conference paper & 2 \\\hline
\end{tabular}
\end{minipage}
\end{table}

The focus on technical aspects must be seen, however, in light of a high number of short papers (cf. Table \ref{tbl:pubtypes}), which due to their page limitations do not provide sufficient space for reporting necessary technical detail. Consequently, the short as well as some of the regular papers may not meet the requirements of the MDE community. The lack of technical detail is also problematic because it restricts replication efforts. It is therefore no surprise that not a single replication study was found (cf. Table \ref{tbl:contribution_types}).


\begin{tcolorbox}[colback=white, colframe=white, boxrule=0mm, sharp corners]
    \textit{MDE approaches that address visual impairments are often abstract and conceptual in nature, do not balance technical and human aspects well, while also lacking technical detail required for assessment and replication.}
\end{tcolorbox}

\begin{tcolorbox}[colback=gray!20, colframe=white, boxrule=0mm, sharp corners]
\textbf{RQ1 Summary:} Research into MDE approaches for visual impairment began late, shows a downward trend, and has produced limited output with low impact in the MDE community. Many of these studies are conceptual, lack technical depth, and are rarely published in high-ranking venues. This highlights a need for more balanced, technically detailed, and widely disseminated research in the field.
\end{tcolorbox}


\subsection{RQ2 -- Addressed visual impairments}
\label{subsec:RQ2}

Within our second research question, we investigate what visual impairments are addressed with model-driven approaches.
This includes investigating both ongoing challenges from the accessibility perspective and the MDE perspective.



\subsubsection{Addressed Visual Impairments}
We investigated what range of visual impairments were addressed in the selected studies (DP 1.1). As pointed out in \autoref{sec:accessibility}, \emph{visual impairments} is a collective term for several conditions with a varying degree of vision loss, and thus different resulting requirements and features in a software. So we wanted to know which visual impairments were addressed and in what context.


\textit{Studies addressing accessibility in general.} 24 out of 30 primary studies address accessibility in general rather than specifically targeting visual impairments (see \autoref{tbl:VisualImpairments}). 
None of these 24 studies provides a clear definition of the visual impairments covered. Most of them 
rely on WCAG or similar standards and guidelines and thus remain at a technical level, opting not to explore socio-technical aspects as a whole. Accordingly, visual impairments are primarily used as examples and not as subjects for detailed study.
%
17 out of 24 studies reference visual impairments as part of their own approach 
most often to provide a representative example of a technical measure. The most frequently used terms used to refer to visual impairments in these 17 studies are still mostly of a generic nature, incl. \emph{visual impairments} (14 times), \emph{blindness} (12 times), \emph{low vision} (7 times), \emph{color blindness} (4 times) and \emph{cataract} (1 time). 
The remaining 7 studies out of 24 do not reference visual impairments as part of their own contributions. 
3 studies discuss visual impairments only in a case study or other form of evaluation, whereas 4 studies only provide examples of visual impairments in the introduction or related work sections.




\begin{table}[ht]
\caption{Addressing visual impairments in our primary studies}
\label{tbl:VisualImpairments}
\footnotesize
\begin{tabular}{|r|l|l|}
\hline
\multicolumn{1}{|l|}{\makecell[l]{\textbf{Number of} \\ \textbf{studies}}} & \textbf{Covering visual impairments} & \textbf{Study Keys}                                        \\ \hline
6   & approach specifically for visually impaired users    &   S1-S3, S6, S10, S17        \\ \hline
17   & \makecell[l]{generic approach but refers to visual impairments \\ in the approach sections} &  \makecell[l]{S4, S5, S8, S11-S13, S18-S22, \\ S24, S25, S27-S30}  \\ \hline
3    & \makecell[l]{generic approach but use a case study/evaluation with \\ visual impairments outside of their own approach sections}     &   S9, S7, S1   \\ \hline
4    & \makecell[l]{generic approach and uses only examples outside of \\ their own approach and case study/evaluation section}  &   S14, S15, S23, S26    \\ \hline
\end{tabular}
\end{table}


\textit{Studies addressing visual impairments.} On the other hand, 6 out of 30 primary studies specifically address visual impairments as expressed through their titles, abstracts and presented novel contributions (see \autoref{tbl:VisualImpairments}). 
Although all 6 studies directly address visual impairments, only 4 studies (S1-S3, S6)
also evaluate visual impairment-related aspects of their approaches via a case study or any other form of evaluation. The most frequently used key words in these 5 studies are: 6x blindness (partial/total) or legally blind; 5x visual impairments or vision disabilities; 5x reduced/low vision or vision loss, 1x color blindness, 1x no light reception, 1x blurred vision. The abstract nature of these keywords indicates that - just like the other 24 studies - these 6 primary studies do not consider the true nature and complexities of visual impairments in detail. Finally only one study (S17)
provides basic definitions for common visual impairments. 


\begin{tcolorbox}[colback=white, colframe=white, boxrule=0mm, sharp corners]
    \textit{The large majority of studies provides generic approaches improving accessibility without studying human-centric aspects of visual impairments in detail.
    We argue in favor of considering accessibility and in the concrete context of this SLR, visual impairments, as first-class citizens and consider them beyond the technical domain when addressing it.}
\end{tcolorbox}


\subsubsection{Target audience}
Another aspect we explore is the target audience of the selected primary studies (DP 1.2). The obvious target audience are, of course, software engineers as the presented approaches were designed to support the integration of accessibility / visual impairment-related needs into MDE approaches and tools. 
22 out of 30 studies  name \emph{software engineers} or \emph{developers} directly (S1-S5, S7-S11, S15-S18, S20, S22-S28), 
while some of these studies also consider other functional roles such as (software) \emph{architect}, \emph{analyst}, \emph{modeler}, \emph{designer},  \emph{service designer}, and \emph{accessibility specialist}. 
In contrast, 4 primary studies refer exclusively to \emph{designers} and \emph{service designers} (S13, S14, S19, S21),
and 4 do not explicitly name their target group (S6, S12, S29, S30).
%
The differentiated view towards involved functional roles can be interpreted to mean that certain specifics need to be taken into consideration to ensure the activities, skills and needs of people acting in these roles are met. Although three studies (S5, S7, S8) 
present evaluations of their MDE tools by software developers, no study explicitly sets out such special considerations. In contrast, studies that do not explicitly state their intended target audience may reflect a technology-centric perspective.

Regarding the evaluation, there is another obvious target audience of the studies included in this SLR. The quality of the artifacts created with an MDE tool - in our case software features that address the accessibility needs of users with visual impairments - can only be effectively validated by end users suffering from these conditions. However, only 6 out of 30 studies present an evaluation with end users (see \autoref{subsec:RQ4} for more details about the evaluation).  
The numbers pick up when looking at studies that explicitly name an end user target group. 17 studies refer to users with visual impairments, while 9 more studies refer more generally to people with disabilities. The latter can be explained by the fact that our search resulted in the inclusion of studies that do not explicitly target the accessibility needs of users with visual impairments but still address visual impairment-related approaches, or present relevant examples. 

\begin{tcolorbox}[colback=white, colframe=white, boxrule=0mm, sharp corners]
    \textit{Most primary studies consider different functional roles to represent anticipated users of their \mde-based approach, and identify beneficiaries of improved accessibility. 
    However, they often overlook or briefly address the specific needs of these audiences, neglecting the human and social aspects of socio-technical systems.
    This shortcoming may hinder understanding of how effectively the tools address accessibility needs or whether they tackle real-world accessibility issues.
}
\end{tcolorbox}


\subsubsection{Accessibility Guidelines}
The accessibility guidelines used in our examined studies (DP 1.3), and the number of studies using them are shown in Table \ref{tbl:guidelines}. 
Most of the studies, namely 21,
used Web Content Accessibility Guidelines (WCAG) international standard developed by The World Wide Web Consortium (W3C), including WCAG 2.0 \cite{WCAG2.0}, WCAG 2.1 \cite{WCAG2.1}, WCAG 2.2 \cite{WCAG2.2}, and WCAG mobile draft \cite{WCAG2.2Mobile}. The second common accessibility guideline used in six of the examined studies is WAI-ARIA \cite{WAI-ARIA1.2},
the Accessible Rich Internet Applications Suite, developed by W3C. Ergonomics of human-system interaction — Part 171: Guidance on software accessibility (ISO 9241-171:2008) \cite{ISO9241}, developed by the International Organization for Standardization (ISO), is the third popular set of guidelines, used by three of the examined studies.
Accessibility development documentation for Android \cite{materialAndroid} and iOS \cite{iosGuide} applications is used in two of the studies. 
There are found to be other guidelines each used in one of the remaining studies, as: Authoring Tool Accessibility Guidelines (ATAG) web standard \cite{ATAG}, developed by W3C; 
BBC mobile accessibility guidelines \cite{BBCMobile} that are based on the requirements of bbc.co.uk content developed for UK audiences and for use with the technology commonly available in the UK; 
IMS guidelines \cite{IMS} for developing accessible learning applications developed by 1EdTech; 
ISO/IEC 12207 \cite{ISO12207} developed by ISO; 
Material Design guideline \cite{materialAndroid} for accessibility developed by Google; 
SIDI guidelines for accessible mobile applications developed by SIDI; 
User Agent Accessibility Guidelines (UAAG) \cite{UAAG} developed by W3C; 
and universal design guidelines \cite{USGovUniversal} by U.S. General Services Administration (GSA). 
Finally, 5 of the studies did not use any guidelines. 

\begin{table}[ht]
\caption{Accessibility guidelines used in our studies}
\label{tbl:guidelines}
\footnotesize
\begin{tabular}{|r|l|l|}
\hline
\multicolumn{1}{|l|}{\makecell[l]{\textbf{Number of} \\ \textbf{Studies}}} & \textbf{Accessibility Guidelines} & \textbf{StudyKeys}                                        \\ \hline
20   & WCAG \cite{WCAG2.0, WCAG2.1, WCAG2.2, WCAG2.2Mobile}   &   \makecell[l]{S2-S5, S7-S11, S13-S15, S18-S20, \\S21, S23-S25, S27, S28}           \\ \hline
6    & WAI / WAI-ARIA \cite{WAI-ARIA1.2}    &   S6, S9, S13, S26, S27, S29      \\ \hline
3    & ISO 9241-171 \cite{ISO9241}    &   S4, S15, S16          \\ \hline
2    & \makecell[l]{Accessibility development documentation \\ for Android \cite{materialAndroid} and iOS \cite{iosGuide}}   &   S2, S9             \\ \hline
1    & ATAG \cite{ATAG}  &   S23                     \\ \hline
1    & BBC mobile accessibility guidelines \cite{BBCMobile} &   S2           \\ \hline
1    & \makecell[l]{IMS guidelines for developing accessible learning \\ applications \cite{IMS}}   &  S28   \\ \hline
1    & ISO/IEC 12207 \cite{ISO12207}   &   S8                            \\ \hline
1     & Material Design guideline for accessibility \cite{materialAndroid}     &   S7      \\ \hline
1     & SIDI guidelines for accessible mobile applications  &  S2    \\ \hline
1     & UAAG \cite{UAAG}           &   S16           \\ \hline
1     & \makecell[l]{Universal design guidelines by U.S. General \\ Services Administration (GSA) \cite{USGovUniversal}} &   S6 \\ \hline
5     & not specified            &   S1, S12, S17, S22, S30    \\ \hline
\end{tabular}
\end{table}

Other guidelines that are mentioned in several studies but have not been used in any of them are: ISO 9126-4 \cite{ISO9126}, A11Y \cite{A11Y}, ISO/IEC 40500:2012 \cite{ISO40500}, BITV, UNE 139803, AODA \cite{AODA}, IEEE Std.610.12, ISO/IEC 9126, European Commission, and national government guidelines, General Accessibility Framework for Administrations (RGAA) \cite{RGAA}, Accessibility for Ontarians with Disabilities Act (AODA), 
Section 508 technical standards \cite{USGovSection508}, initiatives in different countries related to Web accessibility, and UNE 139803.




\begin{tcolorbox}[colback=white, colframe=white, boxrule=0mm, sharp corners]
    \textit{WCAG guidelines are the most used and helpful for developers new to accessibility, but must be tailored to specific project requirements. As they do not focus on particular impairments, end-user design validations and product evaluations are essential to identify missing aspects.}
\end{tcolorbox}

\begin{tcolorbox}[colback=gray!20, colframe=white, boxrule=0mm, sharp corners]
\textbf{RQ2 Summary:} Model-driven approaches for designing and implementing visual impairment needs mainly focus on accessibility challenges rather than MDE-specific issues. Most studies consider functional roles and beneficiaries but provide limited discussion on specific user needs. WCAG guidelines are commonly used but require project-specific adjustments, end-user validations, and product evaluations.
\end{tcolorbox}

\subsection{RQ3 -- Model-Driven Engineering Approaches Used}
\label{subsec:RQ3}

We analyzed which model-driven approaches for the design and implementation of the needs of visually impaired people have been used in our selected primary studies (RQ3). This includes a consideration of which types of applications are developed in the studies, which frameworks and languages the authors use to realize their approaches, and what target application aspects were modeled. 

To better compare how the different approaches work (DP3.3), we have extracted which process steps occur in the primary studies. The chosen process steps are based on typical \mde software engineering processes (see \autoref{fig:approaches}), namely requirements analysis in the analysis phase, modeling (different types of models) and their transformation in the design phase, code generation and manual coding in the implementation phase and testing the resulting application.

\begin{figure}[htb]
    \centering
    \includegraphics[width=\textwidth]{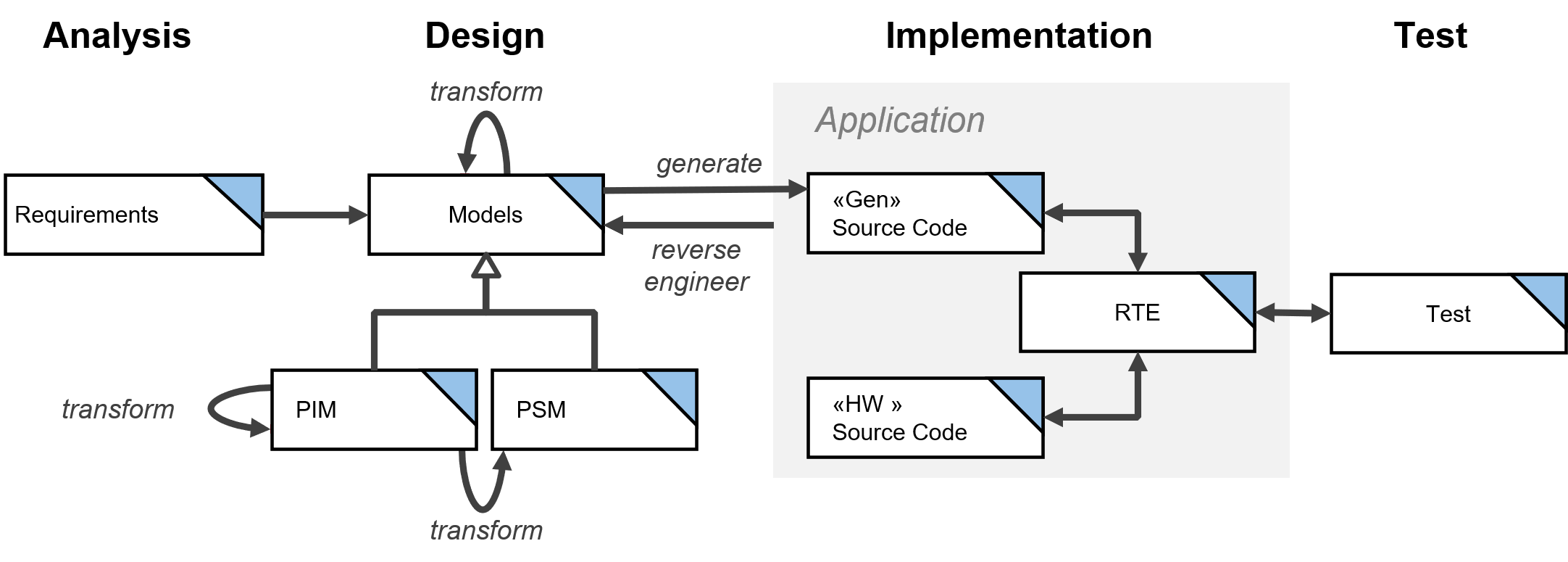}
 \caption{Overview of a generic MDE approach. Steps in which, according to the studies considered, accessibility requirements had to be particularly taken into account by a developer are marked with a triangle in the top right-hand corner.}
      \label{fig:approaches}
\end{figure}

Within \autoref{tab:approach}, we show the results for each primary study regarding the steps in~\autoref{fig:approaches}. In the following, we explain the phases in detail.


\begin{table*}[t]
\caption{Categorization of the approaches based on the phases in the development process.}
\label{tab:approach}
\footnotesize
\centering
\begin{tabular}{|p{0.8cm}||M{1.0cm}|M{0.5cm}|M{0.4cm}|M{0.4cm}|M{1.0cm}|M{0.4cm}|M{0.4cm}|M{0.4cm}|M{0.5cm}|M{0.5cm}| M{0.6cm}|M{0.6cm}|M{1.0cm}|}
\hline
\textbf{Study} & \textbf{Requ.} & \multicolumn{3}{c}{\textbf{Modeling}} & \textbf{M2M} & \multicolumn{3}{c}{\textbf{Generation}}  & \multicolumn{2}{c}{\textbf{Impl.}}  & \multicolumn{2}{c}{\textbf{Test}} & \textbf{Op.}\\
&  & \tiny{GPL DSL} & \tiny{PIM} & \tiny{PSM} & \textbf{Trans.} & \tiny{not spec.} & \tiny{full} & \tiny{core} & \tiny{HWC} & \tiny{Reverse Eng.} &  \tiny{automatic} & \tiny{manual} & \tiny{runtime adaption}\\
\hline
S1  &   &$\bullet$&$\bullet$&  &   &   &  &  &  &  &  & & \\
\hline
S2  &$\bullet$&$\bullet$&  &  &  &$\bullet$&  &  &  &  &  & & \\
\hline
S3 &$\bullet$&$\bullet$&  &  &  &$\bullet$&  &  &  &  &  &  &$\bullet$\\ 
\hline
S4 &  &  &$\bullet$&$\bullet$&$\bullet$&  &  &  &  &  &  & & \\
\hline
S5  &  &$\bullet$&$\bullet$&$\bullet$&$\bullet$&  &$\bullet$&  &  &  &  & & \\
\hline
S6  &  &$\bullet$&  &  &  &  &$\bullet$&  &  &$\bullet$&  & &$\bullet$\\ 
\hline
S7  &  &$\bullet$&  &  &  &  &  &$\bullet$&$\bullet$&  &$\bullet$&$\bullet$& \\
\hline
S8 &$\bullet$&$\bullet$& $\bullet$ &  &  &  &  &$\bullet$&  &  &  & & \\ 
\hline
S9 &  &$\bullet$&  &  &  &$\bullet$&  &  &  &  &  & & \\
\hline
S10  &  &$\bullet$&  &  &  &  &  &$\bullet$&$\bullet$&  &  & & \\
\hline
S11  &$\bullet$&$\bullet$& $\bullet$ &  &  &  &  &  &  &  &  &$\bullet$& \\ 
\hline
S12  &   &$\bullet$&   &  &   &$\bullet$&  &  &  &  &  & & \\
\hline
S13  &$\bullet$&  &$\bullet$&  & $\bullet$ &  &  &  &  &  &  & & \\
\hline
S14  &  &  &$\bullet$&  &  &  &  &  &  &  &  & & \\
\hline
S15  &  &  &$\bullet$&  &  &  &  &  &  &  &  & & \\
\hline
S16  &$\bullet$&$\bullet$&  &  &$\bullet$&  &$\bullet$&  &  &  &  & & \\ 
\hline
S17  &  & $\bullet$ &$\bullet$&$\bullet$&$\bullet$&  &  &$\bullet$&  &  &  & & \\
\hline
S18 &$\bullet$&$\bullet$& $\bullet$ &  &  &  &  &  &  &  &  & & \\ 
\hline
S19  &$\bullet$&$\bullet$&$\bullet$&  &$\bullet$&$\bullet$&  &  &  &  &  & & \\
\hline
S20 &$\bullet$&  &$\bullet$&$\bullet$&$\bullet$&$\bullet$&  &  &  &  &  & &$\bullet$\\ 
\hline
S21 &  &$\bullet$&  &  &  &  &$\bullet$&  &  &  &  &$\bullet$& \\ 
\hline
S22 &$\bullet$&$\bullet$& $\bullet$ &  &  &  &  &  &  &  &  & & \\ 
\hline
S23  &  &$\bullet$& $\bullet$ &  &  &$\bullet$&  &$\bullet$&  &  &  & & \\ 
\hline
S24 &$\bullet$& $\bullet$ &$\bullet$&$\bullet$&$\bullet$&  &  &  &  &  &  & & \\ 
\hline
S25  &$\bullet$&$\bullet$&$\bullet$&$\bullet$&   &   &  &  &  &  &  & & \\
\hline
S26  &$\bullet$&  &$\bullet$&$\bullet$&  &  &  &  &  &  &  & & \\ 
\hline
S27  &  &$\bullet$&$\bullet$&$\bullet$&$\bullet$&$\bullet$&  &  &  &  &  & & \\ 
\hline
S28  &$\bullet$&  &$\bullet$&$\bullet$&$\bullet$&  &  &$\bullet$&  &  &  & & \\ 
\hline
S29  &$\bullet$&$\bullet$&$\bullet$&$\bullet$&  &  &  &$\bullet$&  &  &  & & \\ 
\hline
S30  &$\bullet$&$\bullet$&  &  &  &  &  &  &  &  &  & & \\ 
\hline
\end{tabular}
\end{table*}


\subsubsection{Analysis}

We have investigated which approaches have included accessibility requirements within the analysis phase and how they have included them. 
16 out of 30 studies included requirement analysis elements in their approaches. 
To give some examples, 
S3 has studied and selected 28 accessibility requirements that explicitly target mobile devices. They suggest actionable recommendations when developing native mobile apps and integrate them in a cross-platform framework for \mde
of mobile business apps. 
S2 lists 25 common accessibility requirements in mobile applications which the authors 
compiled from other publications and WCAG 2.1, and describes how they help to avoid certain challenges. They integrate these requirements similar to the work in S3 in extending their \dsl, in the IDE, as well as in changed transformation rules.
S13 lists 5 native requirements for media players and 11 additional requirements to satisfy specific accessibility needs of users. They stay on a very abstract level, e.g., ``Color'', or  ``Find''. They used these requirements when modeling the user interface.  
S18 collects reusable components such as accessibility guidelines, design patterns, and
norms in an accessibility-supportive repository. Requirements are modeled in a graph structure connecting requirements nodes by labeled arcs. However, their connection to the other models or the designed system is not clear and the approach is not replicable. 


\begin{tcolorbox}[colback=white, colframe=white, boxrule=0mm, sharp corners]
\textit{Only slightly more than half of the studies mention the use of accessibility requirements, with few explicitly addressing specific ones. Requirements are often listed but not explicitly modeled, integrated into tools, or manually used to support UI development, limiting reusability for other researchers.}
\end{tcolorbox}

\subsubsection{Modeling} 
We have further investigated, which modeling languages and frameworks were used, categorized the studies if they are using an approach that distinguishes between platform-independent and platform-specific models or not, and analyzed what aspects were modeled. 

\paragraph{Used modeling languages and frameworks} 
To group the \textbf{different modeling approaches}, we have investigated which modeling languages and frameworks were used by the different approaches (DP 3.6).
Most of the studies, namely 23, use \gpls or \dsls (see \autoref{tab:approach}). 

The modeling notation with the most mentions is \uml, namely 10 times (see \autoref{tbl:ModelingLanguages}). 
9 studies 
mention that their approach is based on the Cameleon Reference Framework~\cite{Cameleon03}, a reference for classifying user interfaces supporting multiple contexts. The Cameleon Reference Framework suggests four levels of abstraction: task and concepts, an abstract, concrete, and final user interface. 
These different levels of abstraction can be well handled by \mde  approaches using model-to-model transformations. 
Out of these 9 studies, 5 studies also 
use the User Interface eXtensible Markup Language (UsiXML)~\cite{UsiXML}.  
UsiXML is structured in four levels defined by the Cameleon Reference Framework: Task and Concepts, as well as Abstract, Concrete, and Final User Interfaces.  

6 studies 
use the Eclipse Modeling Framework~\cite{EMF} and 3 out of these studies 
also use the Graphical Modelling Framework (GMF)~\cite{GMF} to define their models. 
5 studies 
reference the \mof structure of instances, models, and meta-models as their underlying principle.  
4 studies 
use UIML~\cite{UIML}, a markup language for user interfaces developed by OASIS Open that provides a canonical XML representation of any user interface. 
4 studies  
mention the EGOKI~\cite{129} system, which generates accessible user interfaces. 
2 of these studies also  
use the SPA4USXML tool~\cite{26}, a part of the EGOKI system which provides functionalities to generate task and abstract UI models (compliant with the USIXML syntax) from a service description,  to relate interaction resources, e.g., texts, images, videos, audios, to abstract interaction elements in abstract UI models, and to generate a resource model.
2 studies
rely on MD$^2$ and Xtext. 
MD$^2$~\cite{MD2} is an approach for the model-driven cross-platform development of mobile applications for Android and iOS. 
Xtext~\cite{Xtext} is a framework enabling the development of programming and domain-specific languages.
One of these studies also mentions using the Xtend~\cite{Xtend} language, a Java dialect and statically typed programming language that translates to Java source code. 
5 studies 
developed its own modeling language, mainly for describing user interfaces
and 4 studies 
lack to mention the used modeling languages or frameworks at all. 

\begin{tcolorbox}[colback=white, colframe=white, boxrule=0mm, sharp corners]
\textit{While one-third of the approaches use a \gpl (\uml) to define models, there is no predominant modeling language, framework, or tool used by the studies. The used languages mainly support the modeling of data or user interfaces.}
\end{tcolorbox}

\begin{table}[ht]
\caption{Used  modeling languages and frameworks}
\label{tbl:ModelingLanguages}
\footnotesize
\begin{tabular}{|r|l|l|}
\hline
\multicolumn{1}{|l|}{\makecell[l]{\textbf{Number of} \\ \textbf{Studies}}} & \textbf{Modeling Languages and Frameworks}  & \textbf{StudyKeys} \\ \hline
10  & UML                      &  S1, S5, S8, S11, S18, S22, S23, S26, S28, S29   \\ \hline
9   & Cameleon Reference Framework  &  S12-S14, S16, S19, S20, S22, S26, S27 \\ \hline
6   & Eclipse Modeling Framework (EMF)   &  S1, S5, S13, S16, S19, S23 \\ \hline
5   & \makecell[l]{User Interface eXtensible Markup \\Language (UsiXML)}  &  S13, S14, S16, S19, S27  \\ \hline
5   & Meta Object Facility (MOF)       &  S5, S14, S19, S23, S28   \\ \hline
4   & UIML   &  S12, S16, S19, S21 \\ \hline
4   & EGOKI   &  S12, S14, S19, S21 \\ \hline
3   & Graphical Modelling Framework (GMF)   &  S13, S16, S19 \\ \hline
2   & SPA4USXML   &  S14, S19 \\ \hline
2   & MD$^2$ and Xtext   &  S2, S3 \\ \hline
1   & Xtend   &  S3 \\ \hline
1   &  ISATINE Framework   &  S4 \\ \hline
1   & WebML    &  S24 \\ \hline
1   & UseML    &  S27 \\ \hline
1   & \makecell[l]{Advanced Adaptation Logic \\Description Language ($AAL\_DL$)}   &  S20 \\ \hline
5  & own DSL developed   &  S7, S9, S10, S17, S22 \\ \hline
4  & no details given   &  S6, S15, S25, S30 \\ \hline
\end{tabular}
\end{table}


\paragraph{Modeled aspects}
We further investigated \textbf{what aspects} of the described applications were \textbf{modeled} (DP3.5, answers see \autoref{tbl:ModeledAspects}).
In summary,
20 studies mentioned that the UI in general or specific screens are modeled and
7 studies referred to the modeling of specific UI components or elements. 
12 studies modeled the interaction between users and the UI or individual actions a user could perform, or a function, or service the user could call. 
8 studies model the appearance, or look and feel	of the UI, and 
7 studies mention the explicit modeling of workflows, navigation, or transition between pages.	
6 studies mention user groups with their specifics and capabilities.
5 studies mentioned the modeling of requirements, or
 the input and the output of services or functions. 	
4 studies used models to define the data structure.  
3 studies described the use of presentation models and
3 studies are modeling context information such as device usage, functions of images, or contextual recommendations.
When looking into the studies in detail, we found it hard to extract more information, e.g., about how accessibility was modeled or providing concrete examples for UI models, including accessibility requirements to be used in generative processes.
There are, however, approaches where one could find more details on modeling the UI (S5, S6), modeling user profiles (S5), or adaptation rules (S13) that can be used as inspiration for developers.

\begin{table}[ht]
\caption{Which aspects were modeled in the selected studies (answers to DP3.5)}
\label{tbl:ModeledAspects}
\footnotesize
\begin{tabular}{|r|l|l|}
\hline
\multicolumn{1}{|l|}{\makecell[l]{\textbf{Number of} \\ \textbf{Studies}}} & \textbf{Modeled aspects}  & \textbf{Studies} \\ \hline
20 & UI general/ screen & S6, S7, S10, S11, S13-S20, S22, S26-S28, S30 \\ \hline 
12 & interaction/ action/ function/ service &  S2, S7, S9-11, S16-S18, S21, S22, S25, S30 \\ \hline 
8 & appearance/ look and feel & S1-S5, S11, S18 S22, \\ \hline
8 &  UI components/ elements &  S1-S5, S7, S24, S25 \\ \hline
7 & workflows/ navigation/ transition &  S4, S9, S10, S23-S25, S30 \\ \hline
6 & user groups, specifics, capabilities & S4, S6, S12, S21, S24, S30 \\ \hline
5 & input/ output & S4, S9, S17, S21, S25  \\ \hline
5 & requirements & S6, S8, S24, S25, S29  \\ \hline
4 & data structure & S8, S16, S24, S25  \\ \hline
3 &  presentation & S4, S24, S30 \\ \hline
3 &  \makecell[l]{ context information \\ (device usage, functions of images, recommendations)}  &  S6, S12, S30 \\ \hline
2 &  adaptation rules & S12, S13 \\ \hline
1 &  information architecture (data, wiki content) & S23  \\ \hline
\end{tabular}
\end{table}

\begin{tcolorbox}[colback=white, colframe=white, boxrule=0mm, sharp corners]
\textit{In our analysis, it was a challenge to identify which models are used in the published applications and which ones are only used for representation in a publication. This was especially hard, as no code repositories were provided.  
Moreover, one has to distinguish between described metamodels and models - a distinction that not all publications explicitly make.
The analysis has shown that there is no main framework or \dsl used. In other words, no standard stands out, nor can it be seen that the trend is moving toward the development of \dsls.  
Surprisingly, none of the \dsls modeled accessibility aspects explicitly, but rather focused on the basic structure of user interfaces. 
This was also shown in analyzing the different aspects modeled: modeling the UI and various details of the UI is a big focus of the approaches. A surprisingly low number of studies model user groups and their needs, relevant context information, or data structures explicitly.
Moreover, it was not possible to extract a suggestion for an accessibility model from the descriptions in these studies as suggestions for other developers, as the presentation was in most cases too abstract and in others too restricted to specific use cases. Thus, the reuse of the concrete modeling approaches is limited.}
\end{tcolorbox}

\subsubsection{Model2Model Transformation} 

10 primary studies mentioned model-to-model transformation in their approaches. 
4 studies (S4, S19, S20, S24) add accessibility aspects to the transformation process starting from general models describing the \ui to models incorporating accessibility requirements.
E.g., S4 adapts their \ui \pims based on information from user disability profiles and the used  UI components to adapted ones, e.g., to use different output devices. This transformation is performed with a set of basic adaptation rules.
Only one study (S5) already started the transformation process with accessible UI models. The authors defined an accessible \ui metamodel and an interaction platform metamodel to reuse this information. 

To further characterize the approaches, 
15 of the primary studies do not specify if their models are platform-dependent or independent. 
10 studies use both platform-independent and platform-specific modeling approaches, to create their applications. However, only 7 out of them provide further details about the model-to-model transformation between \pim and \psm.
5 studies 
only describe a platform-independent modeling approach.


%
%
The other five primary studies transform abstract \ui models to \psms but lack further details. 

\begin{tcolorbox}[colback=white, colframe=white, boxrule=0mm, sharp corners]
\textit{There are only five solutions describing how to incorporate accessibility in models on a more concrete level. While they provide some ideas worth looking into it, these solutions are not reusable for other \mde projects as the detailed accessibility requirements, modeling requirements and adaptation rules are provided by giving some examples only.}
\end{tcolorbox}

\subsubsection{Generation}

\begin{table}[]
\caption{ Model-Driven Approaches in the primary studies}
\label{tbl:Generation}
\footnotesize
\begin{tabular}{|r|l|l|}
\hline
\multicolumn{1}{|l|}{\textbf{Number of Studies}} & \textbf{MDE Impact}  & \textbf{StudyKeys}                                         \\ \hline
12  & Mention MDE               &  S4, S6, S13-S15, S18, S20, S21, S24, S26, S30 \\
7   & Mention Code Generation   &  S2, S3, S9, S11, S19, S23, S27  \\
8   & Generate Part of Application &  S7, S8, S10, S12, S17, S23, S28,  S29, \\
4   & Generate Full Application &  S5, S6, S16, S22 \\
\hline
\end{tabular}
\end{table}

As the next part of an MDE process, we use defined models as input for code generators to synthesize the target code of applications. We have analyzed the studies to identify details of the generation process. The results are presented in \autoref{tbl:Generation}: 12 out of 30 studies (cf. S4, S6, S13-S15, S18, S20, S21, S24, S25, S26, S30) mention MDE; however, the approaches described in the studies stop at the design phase and do not include details of the generative aspects of the MDE process. 7 out of 30 studies (cf. S2, S3, S9, S11, S19, S23, S27) mentioned that they are generating an application without specifying if it is fully running without handwritten additions.
Accessibility is, for example, integrated directly into transformation rules (S2, S3), e.g., by adding and interpreting a focus order.  
%
8 out of 30 studies are generating an application core, a scaffold, or parts of the application (cf. S7, S8, S10, S12, S17, S23, S28, S29).
4 out of 30 studies are generating a full running application, such as an iPhone app for educational content (S5), an android email client (S6), a web interface for a ticket shelter (S22) and an accessible media player (S16) . 

\begin{tcolorbox}[colback=white, colframe=white, boxrule=0mm, sharp corners]
\textit{In summary, many of the studies use MDE in order to generate accessible software. However, few demonstrate their approach with the generation of fully functioning software. No study presents specific adaptations or modifications to the generator that would enable it to produce software with a higher degree of accessibility.}
\end{tcolorbox}

\subsubsection{Implementation}

MDSE approaches in practice have shown that it is important to support handwritten additions to generated code, as there might exist requirements or specific business logic, which can not be generated~\cite{VSB+13}. 
Only two studies mentioned the possibility of adding hand-written additions or additional implementation to the generated code. 
S7 mentions that developers add native functionality for the mobile application which runs together with the generated app project. 
S10 mentions that developers have to add source code to the generated app scaffold which provides required app components, libraries and features such as screen reader support or active voice input.
Moreover, only one study S6 presents a reverse engineering approach, where existing apps on a smartphone are analyzed, and simplified UIs are generated. 

\begin{tcolorbox}[colback=white, colframe=white, boxrule=0mm, sharp corners]
\textit{In summary, we can only conclude limited support for the integration of hand-written implementations, as only two studies explicitly discuss the integration of hand-written artifacts.}
\end{tcolorbox}

\subsubsection{Types of applications developed} 

Despite being often generic and abstract in nature, the vast majority of our selected primary studies explicitly state the type of software environment they focus on (DP 3.2). As shown in~\autoref{tbl:SoftwareType}, most frequently (22 times) mentioned are web applications (S4, S5, S8, S11-S16, S18-S30). 
9 studies (S1-S7, S9, S10) focused on mobile applications. 
Desktop applications were mentioned by two studies (S4, S5), 
and other two studies focus on industrial automation systems (S18, S22). 
1 study (S16) is very specific and discusses accessibility and MDE in the context of Web/HTML5 media players. 
One study (S17) takes a fully generic approach, without committing to a specific application environment. Instead, it provides examples across several contexts, including ATMs, desktop, mobile, and web. S17 is not the only study to span multiple environments: studies S4 and S5 also cover web, mobile, and desktop systems, while studies S18 and S22 focus on web applications alongside industrial automation systems.

\begin{table}[ht]
\caption{Distribution of the application environments covered by our selected primary studies (DP3.2)}
\label{tbl:SoftwareType}
\footnotesize
\begin{tabular}{|r|l|l|}
\hline
\multicolumn{1}{|l|}{\textbf{Number of Studies}} & \textbf{Software Type}  & \textbf{Studies}                                         \\ \hline
22  & Web Application               &  S4, S5, S8, S11-S16, S18-S30   \\
9   & Mobile Application            &  S1-S7, S9, S10 \\
2   & Desktop                       &  S4, S5 \\
2   & Industrial Automation System  &  S18, S22 \\
1   & Media Player                  &  S16 \\
1   & Generic Approach              &  S17\\ \hline
\end{tabular}
\end{table}

\paragraph{Summary and discussion}
The focus on web applications is a natural choice and likely linked to the WCAG accessibility standard most frequently cited by our selected studies which as we know addresses Web accessibility. This makes WCAG's accessibility guidelines directly relevant and at an abstract level straightforward to implement.
The limited amount of attention to mobile applications is surprising and may spell a future work topic. 
Further, considering more than one application environment 
may be explained by the code generation possibilities offered by MDE. 
However, the examples in all four studies are typically not presented in a detail that allows the reader to understand the full MDE `pipeline' from abstract model to executable code. Hence focusing on a single application environment could reduce the amount of information to present and thus increase the clarity and reproducibility of these studies. 

\subsubsection{Test}
Testing is a key aspect of software engineering and application development (cf. \autoref{fig:approaches}). The inclusion of accessibility guidelines should be validated and tested.
3 out of 27 studies (S7, S18, S22) incorporated testing aspects into their approaches. Among them, only 1 study (S8) mentioned using both manual and automated testing while the other 2 focused solely on manual user testing.
(S8) uses an accessibility scanner to find accessibility problems in apps in an automated way together with manual checks of the user interface done by the authors. 

\begin{tcolorbox}[colback=white, colframe=white, boxrule=0mm, sharp corners]
\textit{The low number of approaches showed, that automated testing is not sufficiently considered for MDE approaches for visual impairments. }
\end{tcolorbox}



\subsubsection{Operation}
Next to developing a software that is accessible, we also have to take a look at the operation of such an accessible system. Accessibility can be provided through several means and can be provided through integrated adaptability within the runtime environment.
Among the reviewed studies, only 3 studies integrated run-time adaptation elements.
Khan and Khusro describe in (S6), an approach to model adaptive and thus personalized user interfaces that match the needs of the user.
Similar to Minon et. al who also describe a model-driven adaptive UI approach to meet the needs of the user on an individual level (S11),
Rieger et al. (S3) argue for the need to enable users to adapt an interface according to their needs and show an example; however, no information is provided if this adaptation possibility is modeled or how this is integrated into code generation.

\begin{tcolorbox}[colback=white, colframe=white, boxrule=0mm, sharp corners]
\textit{Only three studies address runtime adaptation, proposing personalized UIs and user-driven interface adjustments, though integration into code generation is rarely detailed. }
\end{tcolorbox}



\begin{tcolorbox}[colback=gray!20, colframe=white, boxrule=0mm, sharp corners]
\textbf{RQ3 Summary:} Model-driven approaches for designing and implementing visual impairment needs show a variety of methods, though explicit modeling of accessibility requirements is not consistently applied. Many studies focus on generating accessible software, primarily targeting web applications due to the relevance of WCAG guidelines. While models often emphasize UI structure, some approaches explore user needs and context-specific aspects. The integration of hand-written artifacts is occasionally addressed, and code generation potential for various application environments is recognized, though detailed examples remain limited. Automated testing is considered in a few cases, highlighting an area with potential for further development.
\end{tcolorbox}

\subsection{RQ4 -- Evaluation} 
\label{subsec:RQ4}
We have further analyzed, how the presented approaches are evaluated. For replicating these studies, access to evaluation data, the code of the developed applications, and the code of the approach leading to these applications is essential. However, none of the studies provided this information.

We have analyzed the evaluation methods and summarized the results. Out of 30 studies, 10 did not include an evaluation, 5 utilized example-based evaluations (see S2, S3, S11, S21, S28), 9 validated their approach through proof-of-concept (refer to S1, S2, S5, S9, S14, S16-S19), and 6 conducted user studies for validation (c.f. S4, S6, S7, S8, S12, S22).

Among these studies, 8 studies provided specific participant numbers in their user studies and proof-of-concept validations: S18 involved 3 software developers in their proof-of-concept, S5 included 4 software developers and 5 students, S4 conducted a study with 42 participants with impairments, S7 involved 42 software developers in their study, S22 executed two studies; the first with 4 software developers and the second with a blind participant, S8 conducted two case studies with students; one with 8 students assessing methodology and application development, and another with 14 students evaluating the MDE tool, and S6 involved 41 blind users using Android smartphones in their study.

For replicating these studies, access to evaluation data, the code of the developed applications, and the code of the approach leading to these applications is essential. However, none of the studies provided this information.


\begin{tcolorbox}[colback=gray!20, colframe=white, boxrule=0mm, sharp corners]
\textbf{RQ4 Summary:} Evaluation methods varied among the reviewed studies: 10 lacked evaluations, 5 used example-based assessments, 9 conducted proof-of-concept validations, and 6 performed user studies. Participant numbers were specified in 8 studies, involving developers, students, and visually impaired users, whereas only 3 studies evaluated their approaches with visually impaired users. However, none of the studies provided access to evaluation data or source code for replication.
\end{tcolorbox}


\subsection{RQ5 -- Reported Strengths and Limitations}
\label{subsec:RQ5}
Our last research question investigates what limitations, gaps, and challenges were reported within the 30 studies.

\subsubsection{Reported limitations in the state-of-practice}

We analyzed what limitations in the state-of-practice motivated the authors of our selected primary studies and what research questions they addressed.
To answer RQ5, we have analysed the abstracts and introductions of each study. If clarity was not fully established in this way, the remaining sections of a study were consulted as well.

Table \ref{tab:motivation} depicts clusters of the main motivational aspects of the reviewed studies. Lack of awareness and attention by developers as well as insufficient tools and methods are the most frequently used aspects to motivate research work. Less frequently used were argumentation lines about low accessibility in existing UI and the potential of adaptable UI to make them more accessible at an individual user level. There are a number of other technical aspects that have been used to motivate reviewed studies as shown in Table \ref{tab:motivation}. Only one \cite{100} of 30 studies took a human-centric approach and motivated the presented research with real-world impact for disadvantaged people.

\begin{table}[ht]
    \centering
    \caption{Number of studies vs main motivational aspect}
    \label{tab:motivation}
    \footnotesize
    \begin{tabular}{|r|l|}
        \hline Frequency & Motivation\\\hline
         7 & Insufficient tools and methods for implementing accessibility \cite{26,27,36,57,79,80, 125}\\        
         6 & Lack of awareness and attention to accessibility needs by developers \cite{3,5,14,17,81,101}\\
         4 & Existing barriers to accessibility in UI \cite{25,29,46,103}\\
         4 & Adaptable UI have the potential to improve individual accessibility \cite{120,72,73,102}\\
         2 & Need for multi-platform support for addressing accessibility needs \cite{6,45}\\
         2 & Low accessibility of mobile apps \cite{8,11}\\
         2 & Extension of Egoki system will simplify the creation of accessible apps \cite{112, 129}\\         
         1 & Wikis lack built-in accessibility support \cite{34}\\
         1 & Lack of accessible Web content \cite{108}\\
         1 & Accessibility is essential for user inclusion and equality \cite{100}\\\hline
    \end{tabular}
\end{table}

The predominant focus on technology to the detriment of human aspects becomes more obvious when looking at the number of studies that directly refer to user needs when motivating their work. Only 2 of 30 studies refer to visual impairments -- the user needs in focus of this SLR. Van Hees and Engelen \cite{29} refer to blind users, while Khan and Khusro \cite{103} use visually impaired and blind users to motivate their work. Neglecting user needs in favour of technology is as worrisome as it is a hindrance when designing solutions aimed at improving human-centric aspects, or more specifically the accessibility, of software systems.

\begin{tcolorbox}[colback=white, colframe=white, boxrule=0mm, sharp corners]
\textit{Most studies are driven by technological limitations, such as developers' lack of awareness or inadequate tools. Human-centric motivations—like real-world impact or addressing specific user needs—are rare. Only one study directly aimed to help disadvantaged users, and just two considered the needs of visually impaired users.}
\end{tcolorbox}

\subsubsection{Reported Strengths of Promising Approaches and Tools} 


In this section, we address the advantages and strengths as reported by the authors of the selected primary studies (DP 4.5).
In the course of our evaluation, several strengths of promising approaches and tools emerged. Firstly, we observed a high frequency of early integration and evaluation of accessibility features, due to the usage of Model-Driven Software Development. The following works S7, S8, S26, S27 and S18 mention that the benefits of model-driven development can be transferred to the development methodologies for accessible applications and yield similar results, such as an "increase in productivity" (S2), "improved development time" and higher flexibility during the development.
Secondly, several studies reported a simplification and of accessibility standard compliance by leveraging Model-Driven Development methods (S2, S7, S17, S18, S24). By creating high-level abstractions of software, these methods provide a comprehensive way to address accessibility requirements while reducing complexity. The separation of problem domains allows the developer to focus on the developed application itself while reducing the required knowledge for the accessibility domain.
Finally, a strength that emerged from the studies was an emphasis on enhancing user efficiency and satisfaction, by using Model-Driven methodologies (S23).
By easing the development of accessibility functionality, the threshold to invest the required resources to do so is lowered thus resulting in a broader implementation of accessibility within the applications.

\begin{tcolorbox}[colback=white, colframe=white, boxrule=0mm, sharp corners]
\textit{The reviewed studies highlight several strengths of Model-Driven Development for accessibility: early integration of accessibility features, improved productivity and flexibility, easier compliance with standards through high-level abstractions, and enhanced user efficiency.}
\end{tcolorbox}

\subsubsection{Reported Limitations} 

This section addresses the reported shortcomings and limitations of the primary studies selected as part of DP 4.5. In general, the number of reported shortcomings and limitations is surprisingly low. Based on the information provided, we have identified eight categories, shown in \autoref{tbl:limits}.
\begin{table}[ht]
\caption{Limitations mentioned in studies.}
\label{tbl:limits}
\footnotesize
\begin{tabular}{|r|l|l|}
\hline
\multicolumn{1}{|l|}{\textbf{Number of Studies}} & \textbf{Mentioned Limitation}  & \textbf{Studies}                                         \\ \hline
6  & Limited Evaluation                            &  S1, S8, S11, S12, S18, S25    \\
4   & Limited Scope                                 &  S18, S25, S29, S30 \\
2   & Limited User Experience                       &  S4, S27 \\
3   & Modeling Difficulties                         &  S14, S23, S24 \\
1   & Too Complex/ Difficult to use                 &  S17 \\
4   & Limited Level of Maturity                     &  S8, S10, S13, S27\\ 
2   & Limited Level adaptabvility                   &  S11, S25\\
2   & Practical Limitations in real-world projects  &  S2, S16 \\ \hline
\end{tabular}
\end{table}
Six articles (S1, S8, S11, S12, S18, S25) reported evaluation limitations. S18 discusses three shortcomings. Firstly, they point out that early-stage testing is difficult because most WCAG criteria are content-related, and thus the testing of early accessibility integration cannot be very comprehensive. Secondly, ``only a subset of accessibility criteria can be tested'' in an automated manner. Manual audit is still required, which impedes potential gains of an MDE approach. Thirdly, their evaluation is focused on ``assistive technologies (AT) for visual impairments'' only. With this, the authors indicate limitations regarding the applicability of their findings to non-visual impairment related accessibility needs. S11 highlights that in both their case studies, only one adaptation rule was used and tested. The support of multiple adaptation rules is earmarked for future work including more evaluations, especially to test potentially conflicting rules. Moreover, more evaluation is required to test incorporated transformations at different abstraction levels. S8 suggest that an automated evaluation has shortcomings by stating that the evaluation by specialists or a test with users, would probably reveal more problems and conclude ``that tests with real users are essential for the identification of other accessibility problems`` These evaluation limitations are of a more serious nature because they limit not only the ability to make statements about the accessibility of created apps but also statements about the reproducibility of presented approaches.

Four studies mention scope limitations or incomplete coverage of accessibility requirements: S18, S25, S29 and S30. S18 highlight that their work was focused on assistive technologies (AT) for visual impairments only, implying the generalizability of their results must be carefully tested. S29 and S30 argue in a similar fashion by stating that basic concepts may be applicable to other accessibility needs than visual impairments but given limitations with supporting interactions and limited multimedia coverage.

S4 and S7 report limited user experiences of their approaches. S4 states that ``while users with high experience with computing platforms were effective with the UI, users with low experience were not as effective.'' S7 explains that their``design is still limited to simple workflows and user interface elements, but can be extended in further versions''. Both issues require further research and user-centered evaluation. 

Modeling difficulties were reported by three studies (S14, S23, S24) incl. the semantics of UI models are hard to generate and need further developer input to provide good results (S14), difficulties to build meta-models that generalize problems (S23), and WebML models that restrict the representation of rich interactions on web pages (S24).

Four studies (S8, S10, S13, S27) report a limited level of maturity and a need for future extensions. S7 explains that the designs that can be generated are ``limited to simple workflows and user interface elements''. S10 reports a need for the ``improvement of [their] concept concerning the utility and usability of the model generation and transformation.'' S13 reports missing tooling (graphical editor) to better support developers. S27 remarks that tool support is only available for Java and PHP programming languages and only considers WCAG 2.0. All the above shortcomings are of either technical or conceptual nature and can be remedied with a reasonable amount of effort.

S2, S16 draw attention to the practical limitations of their research approaches in real-world projects. The essence of their remarks boils down to not just focusing on standard compliance and technological solutions but also considering operational procedures and other such factors. Again these issues require further research and user-centred design and evaluation.

\begin{tcolorbox}[colback=white, colframe=white, boxrule=0mm, sharp corners]
\textit{Few of the 30 reviewed studies report limitations. The most common are limited evaluation and scope, with some noting issues like immature tools, modeling challenges, and restricted user experience. Key concerns include the lack of real-user testing, narrow focus on visual impairments, and low applicability to real-world projects. }
\end{tcolorbox}

\subsubsection{Reported Gaps and Challenges} 

The research gaps of the work under consideration revolve around methodologies for the new implementation and optimization of existing accessibility and the improvement of maintenance in model-driven software development. Only three studies target MDE approaches directly (S23, S24, S28) while the remaining studies have a stronger emphasis on general improvements of accessibility.
\newline
\textbf{(1) Introducing accessibility:} Methodologies that introduce accessibility within an application.
In general, these studies focus on the simplification of the development of accessible desktop applications and mobile apps, by presenting new frameworks, languages, or methodologies. The studies (S4, S14, S17)  focus on the incorporation of accessibility features into the user interface. S26 presents integrated models to introduce accessibility. S9 and S27 focus on the entire architecture of the targeted application in order to introduce accessibility features into the targeted applications, whereas S8 presents a methodology for the integration.
\newline
\textbf{(2) Improving accessibility coverage:} The following studies focus on expanding the accessibility coverage of existing approaches and frameworks (c.f. S2, S3, S5, S6, S7, S10, S11, S13, S15, S28). Studies in this group focus on increasing the number of covered WCAG guidelines of a given approach or reducing the required resources or expertise to meet those guidelines. The studies S11, S13, and S15 discuss adaptation rules to do so. User interfaces are present on a large variety of devices, S3 and S5 focus on cross-platform coverage of accessibility guidelines, and S2, S7, S10 focus especially on mobile applications. The studies S28 and S30 focus on the improvement of guideline coverage for web applications.
\newline
\textbf{(3) Technical / maintenance challenges:} The studies S27, S20 and S21 discuss the benefits of concepts and methodologies for the model-based development of accessible applications. S18 investigates an approach to evaluate accessibility in the very early stages of application development. The study S18 discusses several requirements that should be met during a development process for an application that adheres to accessibility guidelines. In general, these studies claim, that existing techniques do not provide enough support for easy and agile software development of accessible applications. The approach focuses on simplifying the inclusion of accessibility aspects into the targeted software. 
The analyzed studies present the effects of model-driven or at least model-based development on the implementation of accessible applications however, the studies differ in the specific guidelines that were chosen and the amount of guidelines that were covered.

\subsubsection{Open gaps and challenges from accessibility and MDE perspectives} 
We wanted to investigate what specific gaps and open challenges were addressed from an \mde perspective. However, our results showed that the gaps and challenges were rather coming from the accessibility perspective (DP 2.1). 
Out of 30 studies, 27 focused on accessibility as the main research perspective and only used \mde as a method to support this
As a result, these publications do not mention \mde challenges and gaps explicitly.
E.g., 
S8 mentions that web accessibility is not a priority in development projects, that software engineers lack technical knowledge and tools supporting accessibility quickly and simply, and that insufficient resources are allocated for software development projects. This leads the authors to suggest the use of \mde to support the execution of repetitive tasks, reduce the complexity, and improve productivity.
S21 sees gaps rather in ubiquitous system design than in \mde, i.e., that the interfaces downloaded to a user's mobile device are usually the same for all users and often inaccessible for many of them. 

Only 3 studies focus on improving \mde approaches to derive accessible applications. 
S3 explicitly integrates accessibility concerns into the model-driven process of accessible mobile apps and discusses issues that can be automatically handled in the transformations. 
S24 aims to improve the WebML modeling language to become capable of mapping \wcag guidelines to requirements and describes how \mde transformations are derived.
S25 has the aim to develop modeling techniques for handling the non-functional, generic, and crosscutting characteristics of accessibility concerns.

\begin{tcolorbox}[colback=white, colframe=white, boxrule=0mm, sharp corners]
    \textit{The large majority of studies focused on open gaps and challenges from the accessibility perspective, not from the \mde perspective. We see this focus as one of the reasons why none of the primary studies were published at main \mde venues.}
\end{tcolorbox}

\begin{tcolorbox}[colback=gray!20, colframe=white, boxrule=0mm, sharp corners]
\textbf{RQ5 Summary:} The studies on MDE for accessibility highlight strengths such as early integration of accessibility features, improved development productivity, and simplified compliance with accessibility standards through high-level abstractions. Limitations include restricted evaluation due to limited automated testing and content-specific criteria, narrow application scope focused mainly on visual impairments, and technical challenges like complex modeling and insufficient tool support. Identified research gaps involve expanding accessibility standard coverage, improving development methodologies, and addressing maintenance challenges for broader applicability.
\end{tcolorbox}

\section{The Road Ahead}
\label{sec:discussion}

Our analysis of the 30 selected primary studies reveals several shortcomings and areas where further research is needed. While the reviewed studies offer valuable insights and innovative methodologies, certain limitations and ambiguities in their approaches restrict reproducibility and broader applicability to related use cases.  

Despite these constraints, the studies present a wealth of interesting insights and significant potential for advancing the field. By addressing the identified issues and building on existing methodologies, future research can contribute to more robust, transparent, and generalizable solutions. This chapter highlights these opportunities, discusses the limitations, and proposes directions for future work to drive progress in this domain.

\subsection{Identified Limitations}

\subsubsection*{Limitation 1: Measuring Achieved Accessibility} 

Measuring how much of WCAG each study actually implements is extremely difficult because most studies neither specify which conformance level they aim for nor report how many guidelines they meet. Instead, they offer isolated examples—often without any explanation of why those particular checkpoints were chosen—and follow up with thorough, user-based evaluations to validate their work.
In our review, we looked at whether studies covered guidelines broadly or focused deeply on a few (“horizontal versus vertical coverage”), whether they sampled problems by complexity or simply picked examples at random, and how mature their implementations appeared to be. We found that many studies fall into a “vertical prototype” approach, demonstrating one or two recommendations with no rationale for their selection. A few claim full compliance based on automated checkers (e.g. S8,  \url{achecker.ca}), but such tools miss numerous real-world issues, while others fit neither category and leave their actual contributions to accessibility vague and undefined.
The core challenge lies in the very nature of WCAG itself: it abstracts diverse ßhuman needs into technical checkpoints, obscuring the nuances of real-user experience. Automated tools can only catch code-level violations, and many guidelines—such as writing meaningful alt text—demand subjective, manual assessment. Ultimately, ticking off criteria does not guarantee that users can accomplish tasks.

\subsubsection*{Limitation 2: High-level Description of Approaches}

A significant number of studies (S4, S10, S15, S18, S20, S21, S28, S27) indicate that the presented approaches are addressed at a conceptual, abstract or generic level which comes with its own downsides. Most studies do not state their objectives with clearly formulated research questions. Only 4 studies do so ( S2, S5, S7, S18). These omissions can be linked to reduced clarity and reproducibility.

\subsubsection*{Limitation 3: Generalisability and Reproducibility of Approaches}

All analyzed studies present their approaches and methodologies in some form or another. However, not all approaches can be reproduced without further information or access to sources from the authors. In 14 of the studies (S4, S5, S9, S11, S15, S17, S18, S20, S21, S23, S24, S26, S27, 28), the technical details remain unclear. Further, 4 studies (S18, S20, S21, S30) do not state a systematic or detailed discussion. Page and word limits for study submissions are a likely reason for authors to omit this in-depth information in favor of a more detailed description of their findings. The fact that these data are missing makes it almost impossible to reproduce the results or to build on the findings of the work. In addition, although the studies present working methodologies and solutions, they often lack clear reasoning as to why specific steps were taken, making it harder to generalize the presented approach and apply it to similar use cases.

\subsubsection*{Limitation 4: Evaluation with Key Target User Groups}
Few of the inspected studies provided an evaluation of their approaches with real target end users. Of those that did (S1, S8, S11, S12, S18, S25), most had a lack of precisely defined user groups, which made a clear evaluation with a specific target group very difficult. It is often unclear if the presented approach provides a benefit beyond the ones presented in theory. In addition a stronger focus on both sides of stakeholders is often lacking. The studies often either discuss the benefits for the user (e.g. enhanced user experience) or the benefits for the developer (e.g. higher efficiency during development). A simultaneous consideration is missing. 

\subsubsection*{Limitation 5: Lack of MDE Details and Reusability}
Many of the presented studies lack the necessary details to comprehend the implementation in such detail that it can be reproduced. Emphasizing these details and ensuring reusability would facilitate iterative development and continuous improvement. We noticed a lack of specific information on elements such as generators, transformation rules, modeling languages for accessibility (not only user interface description languages) and reusable RTE components. Therefore, we would like to encourage the authors to provide access to repositories and specifications to help advance this field of research.

\subsubsection*{Limitation 6: Accessibility as a Prerequisite Competence}
The presented approaches focus on the inclusion of accessibility guidelines in model-driven software development approaches. A key-aspect of MDE-approaches is its potential of automation. Although the presented approaches include many automation aspects, all of them still rely highly on the developer to provide accessibility expertise. It is very hard to provide full automation in an MDE approach, however there is a high potential to increase the automation of accessibility related tasks in the development process and thus reduce the time and effort spent by developers on building up expertise in accessibility needs and requirements.

\subsubsection*{Limitation 7: Coverage of Accessibility Needs and Guidelines}
The degree in which accessibility needs are covered by a given approach is hard to measure and thus reliable coverage is hard to reach: Very few studies address the specific guidelines that are covered (S1-S3, S6, S10, S17). In addition, the technical perspective often dominates: specific guidelines are often covered as a proof of concept rather than to address a complete scenario for a specific visual impairment. Thus, the effectiveness of the approach is hard to evaluate. However, it should be noted that complete coverage of all guidelines may neither be required nor attainable within a specified use case, as the guidelines can be in conflict with each other. A social perspective, rather than a technical one, offers advantages in this context by directing attention to the guidelines most relevant for a specific visual impairment use case and target users.



\subsection{Research Roadmap}
The field of vision impairment in software development continues to evolve, presenting numerous opportunities to address existing challenges and explore new frontiers. This subsection outlines key areas for future research and innovation (see~\autoref{fig:approaches}) with a focus on vision-related accessibility needs.
%
\begin{figure}[htb]
    \centering
    \includegraphics[width=\textwidth]{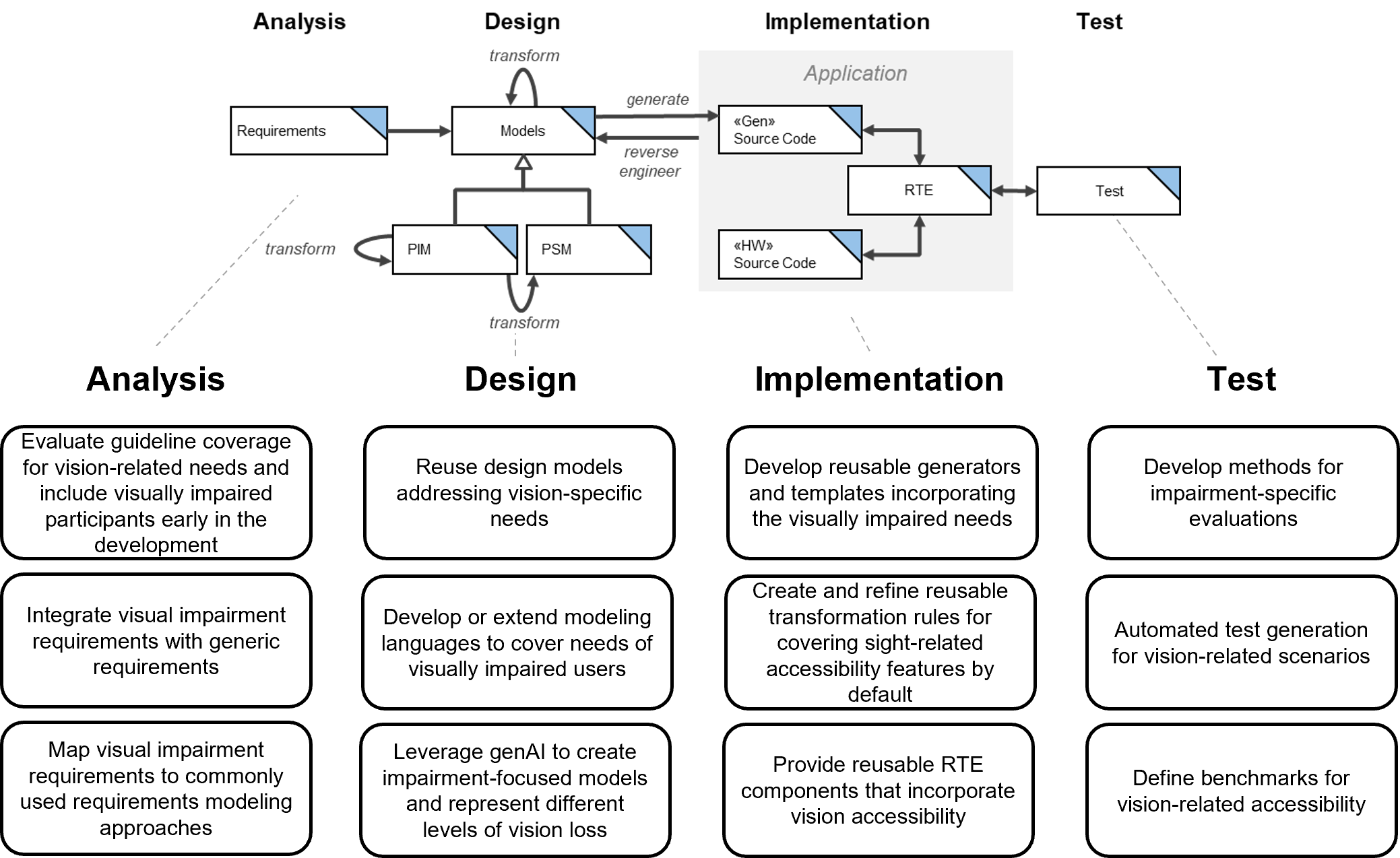}
 \caption{Current gaps and potential areas for further research}
      \label{fig:approaches}
\end{figure}

\subsubsection*{Analysis Phase}
\begin{itemize}
\item \textbf{Evaluate guideline coverage for vision-related needs and include visually impaired participants early in the development.} Investigate the coverage of specific guidelines and standards to ensure comprehensive support for accessibility requirements for different types of visual impairments (e.g., blindness, low vision, color blindness). 
By systematically examining the applicability and thoroughness of these guidelines, we can identify gaps and explore guidelines that are less often covered by research and reveal which impairments are underrepresented and need additional guidance.
When gathering requirements, add user studies with visually impaired participants already early in the development process to refine domain-specific needs beyond standard guidelines.

\item \textbf{Integrate visual impairment requirements with generic requirements.} Explore the integration of specific accessibility requirements for different types of visual impairments with generic requirements for software projects, aiming for seamless alignment. 
The aim is to identify potential conflicts between accessibility guidelines and generic requirements, such as requirements for graphical dashboards vs. needs for non-visual interaction, early in the development process. Early analysis should make it explicit how features translate to voice, tactile or high-contrast modalities.

\item \textbf{Map visual impairment requirements to commonly used requirements modeling approaches.} Provide reusable mappings from impairment-specific needs, e.g., text-to-speech compatibility, zoom support, haptic alters, to  requirements modeling approaches. This aims to bridge the gap between accessibility needs and traditional requirements engineering practices. 
These mappings can facilitate the understanding and systematic incorporation of accessibility in mainstream requirement engineering tools and methods.
\end{itemize}

\subsubsection*{Design and Modeling}
\begin{itemize}
\item \textbf{Reuse design models addressing vision-specific needs.} Investigate how to reuse requirements models that address accessibility needs of users with vision impairments in the design phase to improve efficiency and consistency. 
Reusable models can act as templates (i.e., including constraints for screen reader compatibility, font scaling, contrast ratios or multimodal alternatives such as audio or tactile representations), reducing effort and ensuring that visually impaired users are systematically supported across multiple projects.

\item \textbf{Develop or extend modeling languages to cover needs of visually impaired users.} Evaluate and enhance or develop modeling languages to better represent specific accessibility needs, i.e., annotating color-independent interactions, providing alternative modalities,  tagging semantic description of UI elements, or dependencies on screen readers, in software design. 
By adapting and extending existing modeling languages, developers can more effectively articulate and document accessibility requirements within their designs.

\item \textbf{Leverage genAI to create impairment-focused models and represent different levels of vision loss.} Assess the potential of generative AI tools to define models that effectively address specific accessibility needs, i.e., by configuring prompts or datasets to focus on UI/UX best practices for vision impairments such as voice navigation flows or haptic feedback models, or to propose alternative interaction flows.  
Leveraging AI can help automate and streamline the creation of tailored models, making it easier to incorporate accessibility and provide variants of solutions representing different levels of vision loss.
\end{itemize}

\subsubsection*{Implementation and Generation}
\begin{itemize}
\item \textbf{Develop reusable generators and templates incorporating the visually impaired needs.} Our developed generators and templates should specifically address accessibility needs as defaults, i.e., to produce UIs optimized for screen readers, voice control and magnification tools, facilitating consistent implementation. 
These resources can save time and ensure that accessibility features are implemented uniformly across different projects and teams.

\item \textbf{Create and refine reusable transformation rules for covering sight-related accessibility features by default.}
These rules should enforce, e.g., contrast, scalable texts, alternative text descriptions, color-independent interaction, and keyboard or voice navigation pathways by default and ensure compatibility with assistive technologies such as screen readers or braille displays.
Transformation rules can help automate the process of adapting generic components to meet accessibility standards, reducing manual effort and errors.

\item \textbf{Provide reusable RTE components that incorporate vision accessibility.} Developers should design reusable runtime environment components, such as GUI components, that include settings for specific accessibility needs, e.g., high-contrast themes, text-to-speech toggles, or magnification. 
These components can simplify the development process by providing pre-built, accessible elements that developers can easily integrate into their projects without redesigning accessibility features from scratch.
\end{itemize}

\subsubsection*{Testing}
\begin{itemize}
\item \textbf{Develop methods for impairment-specific evaluations.} We have to develop methods to enable developers to effectively test the level of support and the coverage of accessibility for visually impaired end users. 
This includes creating frameworks and tools to assess how well development tools support accessibility requirements and how accessible the resulting software is for various user groups, e.g., by simulating low-vision conditions, or integrating real screen reader or haptic device testing.

\item \textbf{Automated test generation for vision-related scenarios.} Investigate automated test generation techniques to evaluate accessibility features and ensure compliance with accessibility standards.
This requires to include usage scenarios for visually impaired users, e.g., where missing visual clues are backed by auditory feedback, and to generate test cases that check for missing alternative texts, non-scalable fonts, poor contrast ratios, or inaccessible navigation flows.
Automation can enhance the efficiency and reliability of accessibility testing, allowing developers to identify and address issues more quickly.

\item \textbf{Define benchmarks for vision-related accessibility.} 
The requirements coverage of developed systems should be measured against vision-related accessibility benchmarks, e.g., how well the software functions without visual input or under low-contrast conditions, measuring the proportion of UI elements with semantic annotations, or the availability of auditory or tactile alternatives for visual feedback.

\end{itemize}

Beyond these aspects bound to the \mde process, also other areas could be of interest, e.g., re-engineering approaches to include vision-related accessibility, or simulation approaches for different types and levels of visual impairments.

\section{Threats to Validity}
\label{sec:threats}

Although this study followed well-established best practices for a systematic literature review (\cite{kitchenham2023,wohlin2014}), and great care was taken during its execution, we acknowledge that our research may have been exposed to the following threats.


By following Kitchenham et al. \cite{kitchenham2023} and Wohlin \cite{wohlin2014} for the study design, we minimized threats to internal validity. To ensure all relevant studies were identified, we conducted several trial runs and refinements of our query design to make sure database-specific formats, operator precedence, etc were honored. We also incorporated a systematic snowballing procedure according to Wohlin \cite{wohlin2014} to counter the exclusion of SpringerLink (see Section \ref{sec:datasources} for the reasons of the exclusion). To ensure only studies of suitable quality were considered, we rigorously assessed all studies by (i) removing studies of four or less pages (because space limitations prevent the provision of necessary detail), (ii) filtering out non-peer-reviewed studies, and (iii) carrying out quality assessments during all iterations of the selection procedure. All above activities were assigned to and carried out by a single author, followed by regular team discussions of the results and randomly assigned cross-checks. Threats during study execution were mitigated in a similar fashion: All authors met on a weekly basis during study selection, and data extraction, analysis and synthesis to discuss and verify each others results based on randomly assigned cross-checks. All collected information was stored in a single Cloud-based document. In this way, we minimized threats arising from the manual nature of the majority of these tasks.


To ensure that our selected studies are representative of the state-of-the-art in MDE approaches that address visual impairments, we paired two IT-specific online databases (ACM, IEEE) with two databases that cover a wider range of disciplines (ScienceDirect, Scopus). Further, we carried out forward and backward snowballing to identify studies that did not register hits during the database search. The exclusion of short studies and studies not written in English may affect the representativeness of our set of selected studies. However, we argue that the former are too short to provide sufficient detail, while the latter are rare. English is the SE lingua franca and thus the loss of relevant studies in other languages is unlikely. Finally, we decided not to consider gray literature. Although there is a risk of missing relevant studies, we felt that focusing on peer-reviewed scientific studies represents a good trade-off in favor of the quality of selected studies. 

Another possible limitation are  the keywords included in our search terms (cf Section \ref{sec:searchterm}). To mitigate this threat we included `accessibility' to catch relevant studies that did not use any of the visual impairment keywords in our search term. On the other hand, we omitted MDE-related keywords such as `MDA' or `MDD' from our search term because their inclusion did not result in additional relevant hits during trial, and to keep the search term simple and easy to execute.


\section{Conclusions and Future Work}
\label{sec:conclusions}

While accessibility needs for visually impaired persons are often overlooked in developing socio-technical systems, an \mde approach should provide the needed developer support to incorporate them systematically and consistently throughout the complete application. 
This study presents a systematic literature review of 30 primary studies on the application of model-driven engineering for visual impairments selected from an initial pool of 447 papers. 
We have analyzed existing trends regarding timing, output, impact, and nature of the reported approaches, and what visual impairments they address. We investigated the \mde approaches and their development steps and evaluations in detail, and gave an overview of the reported strengths, limitations, gaps, and challenges.  
Key findings are that most studies to date operate at a high abstraction level, mainly rely on WCAG, and rarely provide reproducible pipelines or working software artifacts.
Only a small number of studies provide concrete modeling approaches for accessibility requirements, functioning implementations, and evaluations with visually impaired end users. This limits reuse, hinders independent verification of the results and constrains the impact for practical adoption. 

Only a few approaches target \mde methods, languages, transformations, code generators, or further tooling as the object of innovation and rather use \mde as an instrument. This might explain why contributions are often rather conceptual, short on implementation details, and underrepresented at core \mde venues. Consequently, the analysis has shown limited research outputs and low visibility. This calls for technically grounded work that treats accessibility as a first-class concern. 
The applied methods are mixed: While WCAG is used by most of the approaches, and models commonly capture the UI, interactions, and navigation, only five studies consider modeling accessibility requirements (without being specific enough to enable reproduction). Evaluations are sparse: 10 studies have no evaluation, and only 6 conducted user studies, whereas only 3 of them involved visually impaired participants. None provided additional evaluation data packages to support replication of the evaluation. Measuring the achieved accessibility of applications remains challenging, as only using automated checkers is not sufficient and conducting user studies is time-intensive.

Using our results, we have sketched some possible research topics in analysis, design, implementation, and testing of accessible applications using MDE. Promising research directions include 
(i) analysis and design techniques that make accessibility requirements explicit, traceable and verifiable at the model level, 
(ii) reusable model transformations, code templates, code generators and runtime components that include accessibility, and
(iii) testing approaches increasing the level of automation and checking for compliance with accessibility requirements. 
Exploring the interplay with methods from AI and generative AI provides additional possible research directions. 
Advancing the field will require publishing more implementation details and replication packages.

In summary, the current state-of-the-art demonstrates potential for improving accessibility of software applications with \mde methods. Providing more reusable artefacts on model, template, and code level and transparent evaluations with relevant user groups, the \mde community can improve development methods for accessibility for visual impairments, delivering an important impact on society.

\begin{acks}
This work is supported in part by ARC Laureate Fellowship FL190100035.
\end{acks}

\bibliographystyle{ACM-Reference-Format}
\bibliography{main}

\appendix

\clearpage
\section{Selected Primary Studies} 
\label{sec:selected-study-appendix}



\begin{small}
\begin{center}
\begin{longtable}{|l|>{\raggedright\arraybackslash}p{2cm}|p{10.7cm}|c|}
\caption{\label{tab:selectedStudies} Overview of selected primary studies} \\
 \hline
    ID & Authors & Title & Year \\\hline
    S1 & Rana et al. & A Novel Model-Driven Approach for Visual Impaired People Assistance OPTIC ALLY \cite{120} & 2022 \\\hline
    S2 & Dias et al. & AccessMDD: An MDD Approach for Generating Accessible Mobile Applications \cite{11} & 2021 \\\hline
    S3 & Rieger et al. & A Model-Driven Approach to Cross-Platform Development of Accessible Business Apps \cite{3} & 2020 \\\hline
    S4 & Bendaly Hlaoui et al. & Model driven approach for adapting user interfaces to the context of accessibility: case of visually impaired users \cite{5} & 2019 \\\hline
    S5 & Bouraoui and Gharbi & Model driven engineering of accessible and multi-platform graphical user interfaces by parameterized model transformations \cite{6} & 2019 \\\hline
    S6 & Khan and Khruso & Blind-friendly user interfaces–a pilot study on improving the accessibility of touchscreen interfaces \cite{103} & 2019 \\\hline
    S7 & Krainz et al. & Can we improve app accessibility with advanced development methods? \cite{8} & 2018 \\\hline
    S8 & Andrade et al. & Incorporating accessibility elements to the software engineering process \cite{100} & 2018 \\\hline
    S9 & Krainz and Miesenberger & Accapto, a generic design and development toolkit for accessible mobile apps \cite{14} & 2017 \\\hline
    S10 & Krainz et al. & Accelerated development for accessible apps – model driven development of transportation apps for visually impaired people \cite{17} & 2016 \\\hline
    S11 & Minon et al. & Integrating adaptation rules for people with special needs in model-based UI development process \cite{72} & 2016 \\\hline
    S12 & Gamecho et al. & Automatic generation of tailored accessible user interfaces for ubiquitous services \cite{129} & 2015 \\\hline
    S13 & Gonz\'{a}lez-Garc\'{\i}a et al. & Adaptation Rules for Accessible Media Player Interface \cite{25} & 2014 \\\hline
    S14 & Mi{\~n}{\'o}n et al. & An approach to the integration of accessibility requirements into a user interface development method \cite{26} & 2014 \\\hline
    S15 & Zouhaier et al. & A MDA-based Approach for Enabling Accessibility Adaptation of User Interface for Disabled People \cite{27} & 2014 \\\hline
    S16 & Gonz\'{a}lez-Garc\'{\i}a et al. & A Model-Based Graphical Editor to Design Accessible Media Players \cite{108} & 2013 \\\hline
    S17 & van Hees and Engelen & Equivalent representations of multimodal user interfaces: Runtime Reification of Abstract User Interface Descriptions \cite{29} & 2013 \\\hline
    S18 & Vieritz et al. & Early accessibility evaluation in web application development \cite{79} & 2013 \\\hline
    S19 & Minon et al. & A graphical tool to create user interface models for ubiquitous interaction satisfying accessibility requirements \cite{112} & 2013 \\\hline
    S20 & Vieritz et al. & User-centered design of accessible web and automation systems \cite{80} & 2011 \\\hline
    S21 & Yazdi et al. & A concept for user-centered development of accessible user interfaces for industrial automation systems and web applications \cite{81} & 2011 \\\hline
    S22 & Abascal et al. & Automatically generating tailored accessible user interfaces for ubiquitous services \cite{73} & 2011 \\\hline
    S23 & Bittar et al. & Accessible Organizational Elements in Wikis with Model-Driven Development \cite{34} & 2010 \\\hline
    S24 & Watanabe et al. & WCAG Conformance Approach Based on Model-Driven Development and WebML \cite{36} & 2010 \\\hline
    S25 & Martin et al. & Engineering accessible Web applications. An aspect-oriented approach \cite{125} & 2010 \\\hline
    S26 & G{\"o}hner et al. & Integrated accessibility models of user interfaces for IT and automation systems \cite{45} & 2008 \\\hline
    S27 & Jeschke et al. & Developing Accessible Applications with User-Centered Architecture \cite{57} & 2008 \\\hline
    S28 & Bouraoui et al. & E-learning and handicap: New trends for accessibility with model driven approach \cite{46} & 2007 \\\hline
    S29 & Vieritz et al. & BeLearning: Accessibility in Virtual Knowledge Spaces \cite{101} & 2007 \\\hline
    S30 & Jeschke and Vieritz & Accessibility and model-based web application development for elearning environments \cite{102} & 2007 \\
  \hline
\end{longtable}
\end{center}
\end{small}

\section{Research Methods}
\label{sec:rm-appendix}
Additional detailed information on selected elements of our research process, as defined in Section \ref{sec:method}.


\begin{small}
\begin{center}
\begin{longtable}{|l|l|l|l|}
\caption{\label{tab:databases} 
Database search parameters} \\
  \hline
    Database & Query & Search options & Scope / SUBJECT AREA \\\hline
    ACM DL & (accessibility OR blind* OR "low vision" OR "vision impair*" & advanced search, & abstract, author keywords\\
    &  OR "visual* impair*") AND (MDE OR "model-driven") & full-text collection & title / - \\
    & & & \\
    IEEE & accessibility OR blind* OR "low vision" OR "vision impair*" & advanced search & abstract, author keywords, \\
    Xplore & OR "visual* impair*" AND MDE OR "model-driven" & & document title / - \\
    & & & \\
    Science & (accessibility OR blind OR "low vision" OR "vision impair" & advanced search, & 'title, abstract or \\
    Direct & OR "visual impair") AND (MDE OR "model-driven") & show all fields & author-specified keywords' / \\
    & & & COMPUTER SCIENCE \\
    & & & \\
    Scopus & (accessibility OR blind* OR \{low vision\} OR "vision impair*" & document search & 'article title, abstract, \\
    & OR "visual* impair*") AND (MDE OR "model-driven") & & keywords' / \\
    & & & COMPUTER SCIENCE \\
  \hline
\end{longtable}
\end{center}
\end{small}


\begin{small}
\begin{center}
\begin{longtable}{|l|l|}
\caption{\label{tab:dataPoints} Overview of our data extraction points} \\
 \hline
    \multicolumn{2}{|l|}{Metadata}\\\hline
    M1 & Publication venue and year\\
    M2 & Number of citations\\
    M3 & Paper type\\
    M4 & Problem type\\
    M5 & Contribution type \\ 
    M6 & Author names and affiliations\\\hline
    \multicolumn{2}{|l|}{Accessibility}\\\hline
    DP1.1 & Addressed visual impairments\\
    DP1.2 & Target audience in focus\\
    DP1.3 & Used accessibility standards and guidelines\\\hline
    \multicolumn{2}{|l|}{Approach and results}\\\hline
    DP2.1 & Addressed gaps and open challenges in the MDE domain [merge DP2.1 and 3.1?]\\
    DP3.1 & Addressed research questions [merge DP2.1 and 3.1?]\\
    DP3.2 & Type of software system in focus\\
    DP3.3 & Technical details of the proposed approach\\
    DP3.5 & Modeled application aspects\\
    DP3.6 & Used MDE frameworks, techniques, models, etc\\
    DP4.1 & Achieved level of MDE process automation\\
    DP4.3 & Achieved level of accessibility\\\hline
    \multicolumn{2}{|l|}{Evaluation, limitations and open challenges}\\\hline
    DP4.4 & Used evaluation method\\
    DP4.5 & Reported pros, cons, limitations and open challenges\\
 \hline
\end{longtable}
\end{center}
\end{small}



\begin{small}
\begin{center}
\begin{longtable}{|l|>{\raggedright\arraybackslash}p{1.6cm}|p{6.6cm}|c|c|c|c|c|c|}
\caption{\label{tab:QA} Quality assessment results} \\
 \hline
    ID & Authors & Title & Year & QA1 & QA2 & QA3 & QA4 & Score \\\hline
    S1 & Rana et al. & A Novel Model-Driven Approach for Visual Impaired People Assistance OPTIC ALLY \cite{120} & 2022 & Y & P & P & N & 2/4 \\\hline
    S2 & Dias et al. & AccessMDD: An MDD Approach for Generating Accessible Mobile Applications \cite{11} & 2021 & Y & Y & P & N & 2.5/4 \\\hline
    S3 & Rieger et al. & A Model-Driven Approach to Cross-Platform Development of Accessible Business Apps \cite{3} & 2020 & P & Y & P & N & 2/4 \\\hline
    S4 & Bendaly Hlaoui et al. & Model driven approach for adapting user interfaces to the context of accessibility: case of visually impaired users \cite{5} & 2019 & Y & Y & P & Y & 3.5/4 \\\hline
    S5 & Bouraoui and Gharbi & Model driven engineering of accessible and multi-platform graphical user interfaces by parameterized model transformations \cite{6} & 2019 & Y & Y & P & N & 2.5/4 \\\hline
    S6 & Khan and Khruso & Blind-friendly user interfaces – a pilot study on improving the accessibility of touchscreen interfaces \cite{103} & 2019 & Y & P & P & N & 2/4 \\\hline
    S7 & Krainz et al. & Can we improve app accessibility with advanced development methods? \cite{8} & 2018 & N & P & N & N & 0.5/4 \\\hline
    S8 & Andrade et al. & Incorporating accessibility elements to the software engineering process \cite{100} & 2018 & P & Y & N & Y & 2.5/4 \\\hline 
    S9 & Krainz and Miesenberger & Accapto, a generic design and development toolkit for accessible mobile apps \cite{14} & 2017 & P & P & N & N & 1/4 \\\hline
    S10 & Krainz et al. & Accelerated development for accessible apps – model driven development of transportation apps for visually impaired people \cite{17} & 2016 & Y & P & N & N & 1.5/4 \\\hline
    S11 & Minon et al. & Integrating adaptation rules for people with special needs in model-based UI development process \cite{72} & 2016 & N & Y & P & N & 1.5/4 \\\hline
    S12 & Gamecho et al. & Automatic generation of tailored accessible user interfaces for ubiquitous services \cite{129} & 2015 & P & P & P & Y & 2.5/4 \\\hline
    S13 & Gonz\'{a}lez-Garc\'{\i}a et al. & Adaptation Rules for Accessible Media Player Interface \cite{25} & 2014 & Y & P & P & N & 2/4 \\\hline
    S14 & Mi{\~n}{\'o}n et al. & An approach to the integration of accessibility requirements into a user interface development method \cite{26} & 2014 & N & Y & P & N & 1.5/4 \\\hline
    S15 & Zouhaier et al. & A MDA-based Approach for Enabling Accessibility Adaptation of User Interface for Disabled People \cite{27} & 2014 & P & P & P & N & 1.5/4 \\\hline
    S16 & Gonz\'{a}lez-Garc\'{\i}a et al. & A Model-Based Graphical Editor to Design Accessible Media Players \cite{108} & 2013 & N & P & P & N & 1/4 \\\hline
    S17 & van Hees and Engelen & Equivalent representations of multimodal user interfaces: Runtime Reification of Abstract User Interface Descriptions \cite{29} & 2013 & Y & Y & P & N & 2.5/4 \\\hline
    S18 & Vieritz et al. & Early accessibility evaluation in web application development \cite{79} & 2013 & N & P & N & N & 0.5/4 \\\hline
    S19 & Minon et al. & A graphical tool to create user interface models for ubiquitous interaction satisfying accessibility requirements \cite{112} & 2013 & N & P & P & N & 1/4 \\\hline
    S20 & Vieritz et al. & User-centered design of accessible web and automation systems \cite{80} & 2011 & P & P & N & N & 1/4 \\\hline
    S21 & Yazdi et al. & A concept for user-centered development of accessible user interfaces for industrial automation systems and web applications \cite{81} & 2011 & N & P & N & N & 0.5/4 \\\hline
    S22 & Abascal et al. & Automatically generating tailored accessible user interfaces for ubiquitous services \cite{73} & 2011 & N & P & N & Y & 1.5/4 \\\hline
    S23 & Bittar et al. & Accessible Organizational Elements in Wikis with Model-Driven Development \cite{34} & 2010 & N & P & P & N & 1/4 \\\hline
    S24 & Watanabe et al. & WCAG Conformance Approach Based on Model-Driven Development and WebML \cite{36} & 2010 & N & P & N & N & 0.5/4 \\\hline
    S25 & Martin et al. & Engineering accessible Web applications. An aspect-oriented approach \cite{125} & 2010 & N & P & P & N & 1/4 \\\hline
    S26 & G{\"o}hner et al. & Integrated accessibility models of user interfaces for IT and automation systems \cite{45}& 2008 & Y & P & N & N & 1.5/4 \\\hline
    S27 & Jeschke et al. & Developing Accessible Applications with User-Centered Architecture \cite{57} & 2008 & P & P & P & N & 1.5/4 \\\hline
    S28 & Bouraoui et al. & E-learning and handicap: New trends for accessibility with model driven approach \cite{46} & 2007 & N & P & P & N & 1/4 \\\hline
    S29 & Vieritz et al. & BeLearning: Accessibility in Virtual Knowledge Spaces \cite{101} & 2007 & N & P & N & N & 0.5/4 \\\hline
    S30 & Jeschke and Vieritz & Accessibility and model-based web application development for elearning environments \cite{102} & 2007 & N & P & N & N & 0.5/4 \\
  \hline
\end{longtable}
\end{center}
\end{small}

\section{Results} 
\label{sec:results-appendix}
Additional detailed information on selected results, as presented in Section \ref{sec:results}.


\begin{small}
\begin{center}
\begin{longtable}{|p{12.4cm}|>{\raggedleft\arraybackslash}p{1.4cm}|}
\caption{\label{tbl:pub_venues} Overview of publication venues} \\
  \hline
    Publication venue & Number of studies\\\hline
    \multicolumn{2}{|l|}{Journals}\\\hline
    Journal on Universal Access in the Information Society & 3 \\\hline
    Journal on Science of Computer Programming & 2 \\\hline
    IEEE Transactions on Human-Machine Systems & 1 \\\hline
    Journal of Multimedia Tools and Applications & 1 \\\hline
    Journal of Universal Computer Science & 1 \\\hline
    Journal on Advances in Human-Computer Interaction & 1 \\\hline
    Journal on Internet and Web Information Systems & 1 \\\hline
    Journal on Multimodal User Interfaces & 1 \\\hline
    Journal on Studies in Health Technology and Informatics & 1 \\\hline
    \multicolumn{2}{|l|}{Conferences}\\\hline
    ACM International Conference on Design of Communication (SIGDOC) & 3 \\\hline
    Innovations in E-learning, Instruction Technology, Assessment, and Engineering Education (ICTA) & 2 \\\hline
    International Conference on Universal Access in Human-Computer Interaction (UAHCI) &	2 \\\hline
    Annual ACM Symposium on Applied Computing (SAC) & 1 \\\hline
    ASEE/IEEE Frontiers in Education Conference & 1 \\\hline
    IEEE/ACIS International Conference on Computer and Information Science (ICIS) & 1 \\\hline
    International ACM SIGACCESS conference on Computers and accessibility (Assets) & 1 \\\hline
    International Conference on Computer Applications in Industry and Engineering (CAINE) & 1 \\\hline
    International Conference on Computers Helping People with Special Needs (ICCHP) & 1 \\\hline
    International Conference on Enterprise Information Systems (ICEIS) &	1 \\\hline
    International Conference on Human-Centred Software Engineering (HCSE), and & 1 \\ 
    International Working Conference on Human Error, Safety, and System Development (HESSDE) & \\\hline
    International Conference on Human Computer Interaction (INTERACCI\'{O}N) & 1 \\\hline
    International Conference on Latest Trends in Electrical Engineering and Computing Technologies (INTELLECT) & 1 \\\hline
    Symposium of the Austrian HCI and Usability Engineering Group & 1\\\hline
    \multicolumn{1}{|r|}{total number of studies} & 30\\
  \hline
\end{longtable}
\end{center}
\end{small}


\begin{small}
\begin{center}
\begin{longtable}{|l|l|>{\raggedleft\arraybackslash}p{1.4cm}|}
\caption{\label{tbl:affiliations} Publications per author team (studies with authors from multiple teams in italic and bold font)}\\
  \hline
    Author team affiliation & Authored studies & Number of studies\\\hline
    University of the Basque Country, Donostia, Spain & \emph{\textbf{S11,}} S12, \emph{\textbf{S14,}} \emph{\textbf{S16, S19,}} S22 & 6 \\\hline
    University of Stuttgart, Germany & \emph{\textbf{S20, S21, S26, S27, S29}} & 5 \\\hline
    ICMC-University of Sao Paulo, Sao Carlos, Brazil & S2, \emph{\textbf{S3,}} \emph{\textbf{S23,}} S24 & 4 \\\hline 
    Technical University Berlin, Germany & \emph{\textbf{S26, S27, S29,}} S30 & 4 \\\hline
    Universidad Carlos III de Madrid, Spain & S13, \emph{\textbf{S14, S16, S19}} & 4 \\\hline 
    FH Joanneum Kapfenberg, Austria & \emph{\textbf{S7, S9,}} S10 & 3 \\\hline 
    RWTH Aachen, Germany & S18, \emph{\textbf{S20, S21}} & 3 \\\hline 
    Johannes Kepler University Linz, Austria & \emph{\textbf{S7, S9}} & 2 \\\hline 
    Laboratory LaTICE, University of Tunis, Tunisia & S4, S15 & 2 \\\hline 
    Ecole Supérieure des Sciences et Techniques de Tunis, Tunisia & S28 & 1 \\\hline
    Federal University of Mato Grosso do Sul (UFMS), Brazil & S8 & 1 \\\hline 
    GULF University for Science and Technology, Mubarak Al-Abdullah, Kuwait & \emph{\textbf{S1}} & 1 \\\hline 
    ISTI-CNR, Pisa, Italy & \emph{\textbf{S11}} & 1 \\\hline 
    Katholieke Universiteit Leuven, Belgium & S17 & 1 \\\hline
    UFPE Recife, Brazil & \emph{\textbf{S23}} & 1 \\\hline 
    UFSCar, Sao Carlos, Brazil & \emph{\textbf{S3}} & 1 \\\hline 
    Universidad Nacional de La Plata and Conicet, La Plata, Argentina & \emph{\textbf{S25}} & 1 \\\hline
    Universidad Nacional del Comahue, Buenos Aires, Argentina & \emph{\textbf{S25}} & 1 \\\hline
    University of Manouba, Tunesia & \emph{\textbf{S5}} & 1 \\\hline 
    University of Muenster, Germany & \emph{\textbf{S3}} & 1 \\\hline 
    University of Peshawar, Peshawar, Pakistan & S6 & 1 \\\hline 
    University of Sialkot, Pakistan & \emph{\textbf{S1}} & 1 \\\hline 
    University of Tunis El Manar, Tunesia & \emph{\textbf{S5}}& 1 \\
  \hline
\end{longtable}
\end{center}
\end{small}

\end{document}